\documentclass[12pt,epsf]{article}
\pdfoutput=1
\usepackage[pdftex]{graphicx}
\usepackage{amsmath,amssymb}
\usepackage{bm}
\usepackage{ascmac}
\usepackage{braket}
\graphicspath{{./fig/}}
\usepackage{comment}
\usepackage{cite}

\newcommand{\beq}{\begin{equation}}
\newcommand{\eeq}{\end{equation}}
\newcommand{\bea}{\begin{eqnarray}}
\newcommand{\eea}{\end{eqnarray}}

\newcommand{\ba}{\begin{aligned}}
\newcommand{\ea}{\end{aligned}}

\def\pe2{p_E^2}

\begin{document}
\setlength{\baselineskip}{0.7cm}
\begin{titlepage} 
\begin{flushright}
OCU-PHYS 527  \\
NITEP 85
\end{flushright}
\vspace*{10mm}%
\begin{center}{\LARGE\bf
Cancellation of One-loop Corrections \\
\vspace*{2mm}
to Scalar Masses in Flux Compactification \\
\vspace*{3mm}
with Higher Dimensional Operators
}
\end{center}
\vspace*{10mm}
\begin{center}
{\Large Takuya Hirose}$^{a}$ and 
{\Large Nobuhito Maru}$^{a,b}$, 
\end{center}
\vspace*{0.2cm}
\begin{center}
${}^{a}${\it 
Department of Mathematics and Physics, Osaka City University, \\ 
Osaka 558-8585, Japan}
\\
${}^{b}${\it Nambu Yoichiro Institute of Theoretical and Experimental Physics (NITEP), \\
Osaka City University, 
Osaka 558-8585, Japan} 
\end{center}
\vspace*{1cm}

\begin{abstract} 
We further study the cancellation of the one-loop corrections to the scalar mass 
 in a six dimensional SU(2) gauge theory with higher dimensional operators, 
 which is compactified on a torus with magnetic flux. 
Higher dimensional operators also contribute to the corrections to the scalar mass nontrivially. 
We explicitly show by the diagrammatic calculations 
 that the corrections are exactly cancelled 
 even with the leading terms of the higher dimensional operators. 
\end{abstract}
\end{titlepage}

\section{Introduction} 
The hierarchy problem has been regarded to be 
 one of the guiding principles to consider the physics 
 beyond the Standard Model (SM) of particle physics. 
In the SM, the quantum corrections to the masses of Higgs scalar field 
 is sensitive to the cutoff scale squared of the theory, 
 which is typically the Planck scale or the scale of the grand unified theory (GUT).  
The experimental values of the Higgs mass 125 GeV requires 
 an unnatural fine-tuning of parameters 
 unless any physical reason such as symmetry reason or dynamical reason is present. 
In order to avoid that fine-tuning, 
 a new physics beyond the SM is believed to be present at order of TeV scale 
 and the quantum corrections to the Higgs mass is of TeV$^2$ order. 
Although supersymmetry or the extra dimensions as a new physics has been considered, 
 there is no such experimental signatures.   
A discrepancy between the scale of new physics and the weak scale 
 is accordingly larger and the degree of unnatural fine-tuning of parameters is enhanced. 
If the quantum corrections to Higgs mass is absent at the new physics scale 
 and is generated at the different scale lower than the new physics scale, 
 Higgs mass can be light even if the new physics scale is at Planck scale, for instance. 

Such a scenario might be possible in the higher dimensional theory 
 with magnetic flux compactification. 
Magnetic flux compactification has been originally studied in string theories \cite{BKLS, IU}. 
Even in the field theory, 
 the flux compactification plays an important role to attempt 
 to explain the number of the generations of the SM fermions \cite{Witten}, 
 to compute Yukawa coupling \cite{CIM, 0903, MS}. 
Recently, it has been paid an attention to the fact 
 that the quantum corrections to the masses for the zero mode of the scalar fields 
 being extra spatial components of the higher dimensional gauge field are cancelled 
 in higher dimensional theory with flux compactification 
 \cite{B1, Lee, B2, HM, 2-loop}\footnote{Similar results are also seen 
 in a six dimensional gauge-Higgs unification compactified on two sphere \cite{LMH}.}. 
The physical reason of cancellation is as follows. 
The translational symmetry in compactified space is spontaneously broken 
 by the vacuum expectation value (VEV) of the extra components 
 of the higher dimensional gauge field. 
In that situation, the zero mode of the scalar fields can be identified with 
 Nambu-Goldstone (NG) boson of the spontaneously broken translational symmetry, 
  which transform as the constant shift and the scalar sector has a shift symmetry. 
This fact implies that only the derivative terms for the scalar fields are allowed in the Lagrangian 
 as in the case of the chiral Lagrangian for pions. 
Therefore, the zero mode of the scalar fields 
 from the extra components of the gauge field in higher dimensions remain massless 
 at the compactification scale.  
If some mechanism breaks the translational symmetry in compactified space explicitly 
 at the different lower scale from the compactification scale, 
 the scalar fields become pseudo NG bosons and can have a light mass 
 as in the case of the pion with a mass 
 due to the explicit chiral symmetry breaking by the quark mass. 
 
In this paper, we further investigate 
 the cancellation of the one-loop corrections to the scalar mass. 
Since the higher dimensional gauge theory is nonrenormalizable, 
 the higher dimensional operators consistent with symmetry of the theory also contribute 
 to the quantum corrections to the scalar mass. 
We will explicitly show in a six dimensional SU(2) Yang-Mills theory with flux compactification 
 that the one-loop corrections to the zero mode of the scalar mass,  
 which are extra components of the higher dimensional gauge field, are indeed cancelled 
 even with the leading terms of the higher dimensional operators. 
The statement is straightforward, 
 but the cancellation is not so trivial since the cancellations of the corrections 
 require the interactions from the operators in different orders.  

This paper is organized as follows. 
In the next section, our setup is introduced. 
The dimensional analysis on the higher dimensional operators is given and 
 the relevant cubic and quartic interactions necessary for the calculations of 
 one-loop corrections to the scalar masses are extracted in section 3. 
Then, the one-loop corrections to the scalar masses are calculated 
 including the leading order terms of higher dimensional operators 
 and are shown to be cancelled in section 4. 
Section 5 is devoted to summary of this paper. 
Details of our calculations are summarized in Appendices.   


\section{Set Up  
}
In this section, we review some results in our previous paper \cite{HM}, 
 which are necessary for our later calculations.

\subsection{Yang-Mills theory with magnetic flux}
We consider a six-dimensional SU(2) Yang-Mills theory with a constant magnetic flux 
 in compactified space. 
The six-dimensional spacetime is represented by $M^4 \times T^2$, 
 where $M^4$ is a four-dimensional Minkowski spacetime 
 and $T^2$ is a two-dimensional torus.
 The Lagrangian of Yang-Mills theory in six dimensions is given by
	\begin{align}
	\mathcal{L}_6 &= -\frac{1}{4} F_{MN}^a F^{aMN}.
	\label{YMlag}
	\end{align}
The field strength tensor is defined by
\begin{align}
	F_{MN}^a = \partial_M A_N^a - \partial_N A_M^a - ig [A_M, A_N]^a, 
\end{align}
where the six-dimensional spacetime indices are denoted by $M, N = 0,1,2, 3, 5, 6$, 
and the gauge indices of SU(2) are denoted by $a,b,c = 1,2,3$. 
$g$ is a SU(2) gauge coupling.  
We follow the metric convention as $\eta_{MN} = (-1, +1, \cdots, +1)$. 

Now, we introduce the magnetic flux in our model.
The magnetic flux is given by the nontrivial background (or VEV) 
 of the fifth and sixth component of the gauge field $A_{5,6}$.
The background of $A_{5,6}$ must satisfy their classical equation of motion $D^m \braket{F_{mn}}=0$. 
Here, $m$ means by extra space indices $m=5,6$ 
and the covariant derivative for the field in the adjoint representation is given by
\begin{align}
D_M A^a_N =\partial_MA^a_N-ig[A_M,A_N]^a
= \partial_M A^a_N + g \varepsilon^{abc} A_M^b A^c_N,  
\end{align}
where $\varepsilon^{abc}$ is a totally anti-symmetric tensor, 
which is a structure constant of SU(2) group. 
Throughout this paper, we choose a solution
\begin{align}
	\braket{A_{5}^{1}}=-\frac{1}{2} f x_{6}, 
	\quad \braket{A_{6}^{1}}=\frac{1}{2} f x_{5}, 
	\quad \braket{A_{5}^{2,3}}=\braket{A_{6}^{2,3}}=0,
	\label{VEVA}
\end{align}
which introduces a constant magnetic flux density $\braket{F^1_{56}}=f$.
Note that this solution breaks a six-dimensional translational invariance spontaneously.
The magnetic flux is obtained by integrating over $T^2$ space and is found to be quantized.
\begin{align}
	\frac{g}{2 \pi} \int_{T^{2}} d x_{5} d x_{6} 
	\left\langle F_{56}^{1}\right\rangle=\frac{g}{2 \pi} L^{2} f=N \in \mathbb{Z},
\end{align}
where $L^2$ is an area of the square torus. For simplicity, we set $L=1$ hereafter.

For convenience, we define $\partial$, $z$, and $\phi$ as
\begin{align}
	\partial \equiv \partial_{z}=\partial_{5}-i \partial_{6}, 
	\quad z \equiv \frac{1}{2} \left(x_{5}+i x_{6}\right), 
	\quad \phi=\frac{1}{\sqrt{2}}\left(A_{6}+i A_{5}\right).
	\label{useful}
\end{align}
In terms of this complex coordinates and variables, 
 the VEV of $\phi$ is given by $\braket{\phi}=f\bar{z}/\sqrt{2}$ from \eqref{VEVA}, 
 and we expand it around the flux background
\begin{align}
	\phi^a=\braket{\phi^a}+\varphi^a,
\end{align}
where $\varphi^a$ are quantum fluctuations of $\phi$ and 
we refer to $\varphi^a$ as scalar fields.

The covariant derivatives in the complex coordinates are useful and defined as
\begin{align}
	DX^a &\equiv (D_5 - iD_6)X^a = \partial X^a-\sqrt{2}g[\phi,X]^a 
	= \mathcal{D}X^a - \sqrt{2}g [\varphi,X]^a, 
	\label{covD} \\
	\bar{D}X^a &\equiv (D_5+iD_6)X^a = \bar{\partial}X^a + \sqrt{2}g[\bar{\phi},X]^a 
	= \bar{\mathcal{D}}X^a+\sqrt{2}g[\bar{\varphi},X]^a, 
	\label{covDbar} \\
	\mathcal{D}X^a &\equiv \partial X^a - \sqrt{2}g [\braket{\phi},X]^a, \\
	\bar{\mathcal{D}}X^a &\equiv \bar{\partial} X^a + \sqrt{2}g [\braket{\bar{\phi}},X]^a,
\end{align}
where $X^a$ means an arbitrary field of the adjoint representation of SU(2). 

Our Lagrangian of SU(2) Yang-Mills theory with gauge-fixing terms and a ghost Lagrangian 
 can be written in a following way. 
	\begin{align}
	\mathcal{L}_{total} = &-\frac{1}{4} F_{\mu\nu}^a F^{a\mu\nu} 
	- \frac{1}{2\xi} D_\mu A^{a\mu} D_\nu A^{a\nu} 
	-\partial_\mu \bar{\phi}^a \partial^\mu \phi^a \nonumber \\
	& -\frac{1}{2} \partial A_\mu^a \bar{\partial} A^{a\mu} 
	+ g^2 [A_\mu, \phi]^a [A^\mu, \bar{\phi}]^a 
	-\frac{g}{\sqrt{2}} \Big\{-\partial A_\mu^a [A^\mu, \bar{\phi}]^a 
	+ \bar{\partial} A^{a\mu} [A_\mu, \phi]^a \Big\} \nonumber \\
	&+ig \Big\{\partial_\mu \phi^a [A^\mu, \bar{\phi}]^a 
	+ \partial^\mu \bar{\phi}^a [A_\mu, \phi]^a \Big\} \nonumber \\
	&-\frac{1}{4} \Big( D \bar{\phi}^a + \bar{D} \phi^a + \sqrt{2}g [\phi, \bar{\phi}]^a \Big)^2 
	+ \frac{\xi}{4} (\mathcal{D} \bar{\phi}^a - \bar{\mathcal{D}} \phi^a)^2 \nonumber \\
	&-\bar{c}^a (D_\mu D^\mu + \xi D_m \mathcal{D}^m) c^a,
	\label{totallag}
	\end{align}
where $\xi$ is a gauge parameter and set to be $\xi=1$ for simplicity. 

\subsection{Kaluza-Klein mass spectrum}
We need to derive mass eigenvalues and eigenstates of the gauge fields and the scalar fields 
 for our calculations.
The covariant derivatives $\mathcal{D}$ and $\bar{\mathcal{D}}$ 
 can be identified with creation and annihilation operators by
\begin{align}
	a = \frac{1}{\sqrt{2gf}} i \bar{\mathcal{D}},~~~
	a^\dag = \frac{1}{\sqrt{2gf}} i \mathcal{D},
	\label{cao}
\end{align}
which satisfy the commutation relation $[a, a^\dag]^{ac} = i \varepsilon^{a1c}$.
Diagonalizing the covariant derivatives, then the commutation relation is diagonalized. 
\begin{align}
	[a,a^\dag] = 
	\left(
	\begin{array}{ccc}
	0 & 0 & 0 \\
	0 & 1 & 0 \\ 
	0 & 0 & -1 \\
	\end{array} 
	\right). 
\end{align}
The mass matrix of gauge fields is accordingly diagonalized as
\begin{align}
	\mathcal{H}^{cc'}_{diag} = -\mathcal{D}_{diag} \bar{\mathcal{D}}_{diag}
	 = 2gf
	\left(
	\begin{array}{ccc}
	n_1 & 0 & 0 \\
	0 & n_2 & 0 \\ 
	0 & 0 & n_3+1 \\
	\end{array} 
	\right),
\end{align}
where $n_{2,3}$ are Landau levels.
The corresponding mass eigenstates of the gauge fields are defined by
\begin{align}
	\widetilde{A}_\mu^a=UA^a_\mu,~~~\widetilde{A}^{a\mu}=U^{-1}A^{a\mu}
	\label{unitaryrot}
	\end{align}
	with a unitary matrix
	\begin{align}
	U=\frac{1}{\sqrt{2}}
	\left(
	\begin{array}{ccc}
	\sqrt{2} & 0 & 0 \\
	0 & 1 & i \\ 
	0 & i & 1 \\
	\end{array} 
	\right).
	\label{unitarymatrix}
\end{align}

The mass matrix of scalar fields is diagonalized as
\begin{align}
	m_{\varphi}^2 = 2gf 
	\left(
	\begin{array}{ccc}
	n_1 & 0 & 0 \\
	0 &  n_2+\frac{1}{2} & 0 \\ 
	0 & 0 & n_3 + \frac{1}{2} \\
	\end{array} 
	\right).
\end{align}
Mass eigenstate of scalar fields are defined by
	 \begin{align}
	\widetilde{\varphi} = U^{-1}\varphi,~~~\overline{\widetilde{\varphi}} = U\bar{\varphi},
	\label{unitaryphi}
	\end{align} 
where $U$ is the same as \eqref{unitarymatrix}.

\subsection{Kaluza-Klein expansion and mode function}
By diagonalizing the covariant derivatives \eqref{cao}, 
 each component of creation and annihilation operators are obtained as follows. 
	\begin{align}
	\begin{cases}
	a_1\equiv\dfrac{1}{\sqrt{2gf}}i\bar{\partial} \\\\
	a_2\equiv\dfrac{1}{\sqrt{2gf}}i(\bar{\partial} + gfz) \\\\
	a_3\equiv\dfrac{1}{\sqrt{2gf}}i(\bar{\partial} - gfz)
	\end{cases}
	, \qquad
	\begin{cases}
	a_1^\dag\equiv\dfrac{1}{\sqrt{2gf}}i\partial \\\\
	a_2^\dag\equiv\dfrac{1}{\sqrt{2gf}}i(\partial - gf\bar{z}) \\\\
	a_3^\dag\equiv\dfrac{1}{\sqrt{2gf}}i(\partial + gf\bar{z})
	\end{cases}.
	\label{creation}
	\end{align}
Note that $a_1$ and $a_1^\dag$ play no role of creation and annihilation operators.  
Although $a_2$ and $a_2^\dag$ are ordinary annihilation and creation operators, 
 the roles of creation and annihilation operators for $a_3$ and $a_3^\dag$ are inverted because of $[a_3, a_3^\dag]=-1$. 
By using \eqref{creation}, non-zero mode eigenfunctions $\psi^a_{n_a,j}$ are constructed 
 similar to the harmonic oscillator, 
	\begin{align}
	\psi^1_{n_1,j},~~~
	\psi^2_{n_2,j} = \frac{1}{\sqrt{n_2!}} (a_2^\dag)^{n_2} \psi^2_{0,j},~~~
	\psi^3_{n_3,j} = \frac{1}{\sqrt{n_3!}} (a_3)^{n_3} \psi^3_{0,j},
	\end{align}
and satisfy orthonormality conditions
	\begin{align}
	\int_{T^2} dx^2 (\psi_{n_a',j'}^{a'})^* \psi_{n_a,j}^a
	= \delta^{a'a} \delta_{n_a'n_a} \delta_{j'j}.
	\label{ortho}
	\end{align}

To derive a four dimensional effective Lagrangian by KK reduction, 
 we need to expand $A^a_\mu$ and $\varphi^{2,3}$ in terms of mass eigenfunctions $\psi^a_{n_a,j}$
	\begin{align}
	\widetilde{A}_\mu^a &= \sum_{n_a,j}\widetilde{A}_{\mu,n_a,j}^a\psi_{n_a,j}^a~(a=1,2,3), \\
	\widetilde{\varphi}^{a} &= \sum_{n_a,j} \widetilde{\varphi}^a_{n_a,j}\psi^a_{n_a,j},~~~\overline{\widetilde{\varphi}}^a=\sum_{n_a,j}\overline{\widetilde{\varphi}}^a_{n_a,j}\psi^{a*}_{n_a,j}~(a=2,3).
	\end{align}

\subsection{Effective Lagrangian}
By substituting these mode expansions into Lagrangian \eqref{totallag} 
 and performing a integration on $T^2$, the four dimensional effective Lagrangian can be obtained. 
Here, we list only cubic terms with a single $\varphi^1$ or $\overline{\varphi}^1$ 
 which are required for our later calculations.\footnote{The sign is different form the results in \cite{HM}, 
but we verified that the sign in the present paper is correct.}
\begin{align}
	\mathcal{L}_{\varphi A A}&=+\sum_{n, j} \frac{g \sqrt{\alpha\left(n+1\right)}}{\sqrt{2} i} 
	\widetilde{A}_{\mu, n, j}^{2} \widetilde{A}_{n+1,j}^{2 \mu} \bar{\varphi}^{1}-\sum_{n, j} 
	\frac{g \sqrt{\alpha\left(n+1\right)}}{\sqrt{2} i} \widetilde{A}_{\mu, n, j}^{3} \widetilde{A}_{n+1,j}^{3 \mu} \bar{\varphi}^{1} \nonumber \\
	&\quad-\sum_{n, j} \frac{g \sqrt{\alpha\left(n+1\right)}}{\sqrt{2} i} 
	\widetilde{A}_{\mu, n, j}^{2} \widetilde{A}_{n+1,j}^{2 \mu} \varphi^{1}+\sum_{n, j} 
	\frac{g \sqrt{\alpha\left(n+1\right)}}{\sqrt{2} i} \widetilde{A}_{\mu, n, j}^{3} \widetilde{A}_{n+1,j}^{3 \mu} \varphi^{1}, 
	\label{paperphiAA} \\
	\mathcal{L}_{\bar{\varphi} \bar{\varphi} \varphi}&=+\sum_{n, j} 
	\frac{g \sqrt{\alpha\left(n+1\right)}}{\sqrt{2} i} \overline{\widetilde{\varphi}}_{n+1, j}^{2} \widetilde{\varphi}_{n, j}^{2} 
	\bar{\varphi}^{1}-\sum_{n, j} \frac{g \sqrt{\alpha\left(n+1\right)}}{\sqrt{2} i} \overline{\widetilde{\varphi}}_{n, j}^{3} 
	\widetilde{\varphi}_{n+1, j}^{3} \bar{\varphi} \nonumber \\
	&\quad-\sum_{n, j} \frac{g \sqrt{\alpha\left(n+1\right)}}{\sqrt{2} i} \widetilde{\varphi}_{n+1, j}^{2} 
	\overline{\widetilde{\varphi}}_{n, j}^{2} \varphi^{1}+\sum_{n, j} \frac{g \sqrt{\alpha\left(n+1\right)}}{\sqrt{2} i} 
	\widetilde{\varphi}_{n, j}^{3} \overline{\widetilde{\varphi}}_{n+1, j}^{3} \varphi^{1}, 
	\label{paperphiphi}
\end{align}
where $\alpha=2gf$. 

\section{Higher Dimensional Operator}
Since the higher dimensional gauge theory is nonrenormalizable, 
 the higher dimensional operators which are consistent with symmetry of the theory should be considered. 
The main purpose of this paper is to show that 
 one-loop quantum corrections to the masses of scalar fields $\varphi^1, \bar{\varphi}^1$ are cancelled 
 even if we take into account the contributions from the higher dimensional operators. 
Before going to the calculation in detail, we classify the higher dimensional operators 
 based on a dimensional analysis. 

In general, we can add the gauge invariant higher dimensional operators to Lagrangian \eqref{YMlag}. 
	\begin{align}
	\mathcal{L}_6 = -\frac{1}{4}F^a_{MN}F^{aMN} + \frac{1}{\Lambda^2} \mathcal{O}_1(D,F) 
	+ \frac{1}{\Lambda^4} \mathcal{O}_2(D,F) + \frac{1}{\Lambda^6} \mathcal{O}_3(D,F) + \cdots,
	\end{align}
where $\mathcal{O}_n(D,F)$ is a set of gauge invariant operators with covariant derivatives and field strengths. 
$\Lambda$ is a cutoff scale of the theory and $n$ is an order of $1/\Lambda^2$.
For $\mathcal{O}_n(D,F)$, 
we can determine the form of operators allowed in $\mathcal{O}_n(D,F)$ 
 by considering mass dimension in four dimension of $\mathcal{O}_n(D,F)$.
In the case of $n=1$ (the first order in $1/\Lambda^2$), 
 there are three allowed operators.
	\begin{align}
	\mathcal{O}_1(D,F)=D^4F+D^2F^2+F^3.
	\end{align}
Similarly, in the case of $n=2$ (the second order in $1/\Lambda^2$), 
 there are four allowed operators. 
	\begin{align}
	\mathcal{O}_2(D,F)=D^6F+D^4F^2+D^2F^3+F^4. 
	\end{align}
More explicitly, the operators $\mathcal{O}_1(D,F)$ and $\mathcal{O}_2(D,F)$ are written 
 by\footnote{For convenience of the calculation, a factor ``2" is included in the second term of $\mathcal{O}_1(D, F)$ 
to cancel a factor 1/2 coming from the normalization condition for the generators.}
	\begin{align}
	\mathcal{O}_1(D,F)&=\mathrm{Tr}[D_LD^LD_MD_NF^{MN}]+2\mathrm{Tr}[D_LF_{MN}D^LF^{MN}] 
	\nonumber \\
	&\quad +\epsilon^{M_1 N_1 M_2 N_2 M_3 N_3}\mathrm{Tr}[F_{M_1 N_1}F_{M_2 N_2}F_{M_3 N_3}], 
	\label{O1}\\
	\mathcal{O}_2(D,F)&=\mathrm{Tr}[D_KD^KD_LD^LD_MD_NF^{MN}]+\mathrm{Tr}[D_kD_LF_{MN}D^KD^LF^{MN}] \nonumber \\
	&\quad +\mathrm{Tr}[\epsilon^{M_1N_1M_2N_2M_3N_3}(D_LF_{M_1N_2})(D^LF_{M_2N_2})F_{M_3N_3}]
	+\mathrm{Tr}[F_{MN}F^{MN}F_{AB}F^{AB}], \label{O2}
	\end{align}
where $\epsilon^{M_1 N_1 M_2 N_2 M_3 N_3}$ is a totally anti-symmetric tensor.


In this paper, we mainly focus on the operators \eqref{O1}, 
 which are the leading terms of the higher dimensional operators. 
Since only the second term in \eqref{O1} will be found to be non-vanishing, 
 we derive the cubic terms with a single $\varphi^1$ or $\overline{\varphi}^1$ and 
 the quartic terms involving two $\varphi^1$ and $\overline{\varphi}^1$ from it, 
 which are necessary for calculations of one-loop corrections to scalar mass.

\subsection{$\mathrm{Tr}[D_LD^LD_MD_NF^{MN}]$} \label{DDDF}
This operator vanishes because of the traceless condition for  SU(2) generators. 
\begin{align}
	\mathrm{Tr}[D_LD^LD_MD_NF^{MN}]=(D_LD^LD_MD_NF^{MN})^a\mathrm{Tr}[t^a]=0.
\end{align}
Thus, we need not calculate the first term in \eqref{O1}.

\subsection{$\epsilon^{M_1 N_1 M_2 N_2 M_3 N_3}\mathrm{Tr}[F_{M_1 N_1}F_{M_2 N_2}F_{M_3 N_3}]$} \label{FFF}
The third term in \eqref{O1} also vanishes 
 because of properties of totally anti-symmetric tensor $\epsilon^{M_1 N_1 M_2 N_2 M_3 N_3}$ 
 and the trace of three generators.
We first note that the trace of $t^a t^b t^c$ is written as
\begin{align}
	\mathrm{Tr}\left[t^{a} t^{b} t^{c}\right]
	&=\frac{1}{4} i \epsilon^{a b c}.
\end{align}
Using this result, we can find the third term to take the following form. 
\begin{align}
	\epsilon^{M_1 N_1 M_2 N_2 M_3 N_3}\mathrm{Tr}[F_{M_1 N_1}F_{M_2 N_2}F_{M_3 N_3}]
	&=\frac{i}{4}\epsilon^{a b c}\epsilon^{M_1 N_1 M_2 N_2 M_3 N_3}F^a_{M_1 N_1}F^b_{M_2 N_2}F^c_{M_3 N_3} 
	\nonumber \\
	&=-\frac{i}{4}\epsilon^{a b c}\epsilon^{M_1 N_1 M_2 N_2 M_3 N_3}F^a_{M_1 N_1}F^b_{M_2 N_2}F^c_{M_3 N_3},
\end{align}
where we interchanged the indices $a \leftrightarrow b$ and $M_1,N_1 \leftrightarrow M_2,N_2$ in the second equality, 
and used the properties of two anti-symmetric tensors $\epsilon^{abc}$ and $\epsilon^{M_1 N_1 M_2 N_2 M_3 N_3}$.
Then we conclude 
\begin{align}
	\epsilon^{M_1 N_1 M_2 N_2 M_3 N_3}\mathrm{Tr}[F_{M_1 N_1}F_{M_2 N_2}F_{M_3 N_3}]=0. 
\end{align}

\subsection{$2\mathrm{Tr}[D_LF_{MN}D^LF^{MN}]$}
Finally, we discuss the second term in \eqref{O1}. 
This operator can be decomposed into the fields 
 with four-dimensional and extra two-dimensional indices as follows.
\begin{align}
	2\mathrm{Tr}[D_LF_{MN}D^LF^{MN}]&=D_LF^a_{MN}D^LF^{aMN} \nonumber \\
	&=D_\rho F^a_{\mu\nu}D^\rho F^{a\mu\nu}+2D_\rho F^a_{\mu m}D^\rho F^{a\mu m}
	+2D_\rho F^a_{56}D^\rho F^{a56} \nonumber \\
	&\quad\quad+D_l F^a_{\mu\nu}D^l F^{a\mu\nu}
	+2D_l F^a_{\mu m}D^l F^{a\mu m}+2D_l F^a_{56}D^l F^{a56}.
	\label{DFDF}
\end{align}
Since the first term does not have terms with $\varphi^1, \bar{\varphi}^1$, 
 it is irrelevant to our calculations. 
We then decompose the remaining terms in \eqref{DFDF} in more detail.
\begin{align}
	2D_\rho F^a_{\mu m}D^\rho F^{a\mu m}
	&\supset
	+4g\varepsilon_{abc}\partial_\mu\partial_\nu\varphi^a\partial^\nu A^{b\mu}\bar{\varphi}^c
	+4g\varepsilon_{abc}\partial_\mu\partial_\nu\bar{\varphi}^a\partial^\nu A^{b\mu}\varphi^c 
	\nonumber \\
	&\quad-2\sqrt{2}ig\varepsilon_{abc}\partial_\mu\mathcal{D} A^a_\nu\partial^\mu A^{b\nu}\bar{\varphi}^c
	+2\sqrt{2}ig\varepsilon_{abc}\partial_\mu\bar{\mathcal{D}} A^a_\nu\partial^\mu A^{b\nu}\varphi^c 
	\nonumber \\
	&\quad+4g^2\bar{\varphi}^{a}\varphi^a\partial_\mu A^b_\nu\partial^\mu A^{b\nu}
	-4g^2\bar{\varphi}^{a}\varphi^b\partial_\mu A^a_\nu\partial^\mu A^{b\nu},
	\label{2ndterm}
\end{align}
\begin{align}
	2D_\mu F_{56}^aD^\mu F^{a56}
	&\supset -2\sqrt{2}g\partial_\mu(\mathcal{D}\bar{\varphi}^a
	+\bar{\mathcal{D}}\varphi^a)[\partial^\mu\varphi,\bar{\varphi}]^a
	-2\sqrt{2}g\partial_\mu(\mathcal{D}\bar{\varphi}^a
	+\bar{\mathcal{D}}\varphi^a)[\varphi,\partial^\mu\bar{\varphi}]^a  
	\nonumber \\
	&\quad +2g^2[\partial_\mu\varphi,\bar{\varphi}]^a[\partial^\mu \varphi, \bar{\varphi}]^a
	+4g^2[\partial_\mu\varphi,\bar{\varphi}]^a[\varphi,\partial^\mu\bar{\varphi}]^a
	+2g^2[\varphi,\partial_\mu\bar{\varphi}]^a[\varphi,\partial^\mu\bar{\varphi}]^a  
	\nonumber \\
	&\quad -2\varepsilon_{abc}\partial_\mu(\mathcal{D}\bar{\varphi}^a
	+\bar{\mathcal{D}}\varphi^a)(\mathcal{D}\bar{\varphi}^b
	+\bar{\mathcal{D}}\varphi^b)A^{c\mu}   \nonumber \\
	&\quad +g^2A^a_{\mu}A^{a\mu}(\mathcal{D}\bar{\varphi}^b
	+\bar{\mathcal{D}}\varphi^b)(\mathcal{D}\bar{\varphi}^b+\bar{\mathcal{D}}\varphi^b) 
	\nonumber \\
	&\quad -g^2A^a_{\mu}A^{b\mu}(\mathcal{D}\bar{\varphi}^a+\bar{\mathcal{D}}\varphi^a)(\mathcal{D}\bar{\varphi}^b
	+\bar{\mathcal{D}}\varphi^b),
	\label{3rdterm}
\end{align}
\begin{align}
	D_l F^a_{\mu\nu}D^l F^{a\mu\nu}
	&\supset2\sqrt{2}ig\varepsilon_{abc}\varphi^a\Big(\partial_\mu \bar{\mathcal{D}} A_\nu^b\partial^\mu A^{c\nu}
	-\partial_\mu \bar{\mathcal{D}} A_\nu^b\partial^\nu A^{c\mu}\Big) 
	\nonumber \\
	&\quad-2\sqrt{2}ig\varepsilon_{abc}\bar{\varphi}^a\Big(\partial_\mu 
	\mathcal{D} A_\nu^b\partial^\mu A^{c\nu}-\partial_\mu \mathcal{D} A_\nu^b\partial^\nu A^{c\mu}\Big) 
	\nonumber \\
	&\quad+4g^2\bar{\varphi}^a\varphi^{a}(\partial_\mu A^a_\nu\partial^\mu A^{a\nu}
	-\partial_\mu A^a_\nu\partial^\nu A^{a\mu}) \nonumber \\
	&\quad-4g^2\bar{\varphi}^a\varphi^b(\partial_\mu A^a_\nu\partial^\mu A^{b\nu}
	-\partial_\mu A^a_\nu\partial^\nu A^{b\mu}),
	\label{4thterm}
\end{align}
\begin{align}
	2D_l F^a_{\mu m}D^l F^{a\mu m}
	&\supset-2\sqrt{2}g\bar{\mathcal{D}}(\partial_\mu\varphi^a)[\varphi,\partial^\mu\bar{\varphi}]^a
	+2\sqrt{2}g\mathcal{D}(\partial_\mu\bar{\varphi})[\bar{\varphi},\partial^\mu\varphi]^a 
	\nonumber \\
	 & \quad -4g^2[\bar{\varphi},\partial^\mu\varphi]^a[\varphi,\partial^\mu\bar{\varphi}]^a 
	-4ig\bar{\mathcal{D}}(\partial_\mu\varphi^a)[\varphi,\bar{\mathcal{D}} A^\mu]^a
	+4ig\mathcal{D}\bar{\mathcal{D}} A^a_\mu[\bar{\varphi},\partial^\mu\varphi]^a 
	\nonumber \\
	&\quad -4ig\bar{\mathcal{D}}(\partial_\mu\varphi^{a})\mathcal{D}([A^{\mu},\bar{\varphi}]^a) 
	+4ig\bar{\mathcal{D}}\mathcal{D} A^a_\mu[\varphi,\partial^\mu\bar{\varphi}]^a
	-4ig[\bar{\varphi},\mathcal{D} A_\mu]^a\mathcal{D}(\partial^\mu\bar{\varphi}^a)
	\nonumber \\
	&\quad -4ig\bar{\mathcal{D}}([A_\mu,\varphi]^a)\mathcal{D}(\partial^\mu\bar{\varphi}^a) 
	-2\sqrt{2}g\bar{\mathcal{D}}\mathcal{D} A^a_\mu[\varphi,\bar{\mathcal{D}} A^\mu]^{a}
	+2\sqrt{2}g[\bar{\varphi},\mathcal{D} A_\mu]^{a}\mathcal{D}\bar{\mathcal{D}} A^{a\mu}
	\nonumber \\
	&\quad -4g^2[\bar{\varphi},\mathcal{D} A_\mu]^{a}[\varphi,\bar{\mathcal{D}} A^\mu]^{a} 
	-2\sqrt{2}g\bar{\mathcal{D}}\mathcal{D} A^a_\mu\mathcal{D}([A^\mu,\bar{\varphi}]^a)
	+4g^2\bar{\mathcal{D}}\mathcal{D} A^a_\mu[\varphi,[A^\mu,\bar{\varphi}]]^{a}
	\nonumber \\
	&\quad -4g^2[\bar{\varphi},\mathcal{D} A_\mu]^{a}\mathcal{D}([A^\mu,\bar{\varphi}]^a) 
	+2\sqrt{2}g\bar{\mathcal{D}}([A_\mu,\varphi]^a)\mathcal{D}\bar{\mathcal{D}} A^{a\mu}
	\nonumber \\
	&\quad -4g^2\bar{\mathcal{D}}([A_\mu,\varphi]^a)[\varphi,\bar{\mathcal{D}} A^\mu]^{a}
	+4g^2[\bar{\varphi},[A_\mu,\varphi]]^{a}\mathcal{D}\bar{\mathcal{D}} A^{a\mu} 
	\nonumber \\
	&\quad-4g^2\bar{\mathcal{D}}([A_\mu,\varphi]^a)\mathcal{D}([A^\mu,\bar{\varphi}]^a),
	\label{5thterm}
\end{align}
\begin{align}
	2D_l F^a_{56}D^l F^{a56}
	&\supset-\sqrt{2}g\bar{\mathcal{D}}\mathcal{D}\bar{\varphi}^a\mathcal{D}([\varphi,\bar{\varphi}]^a)
	-\sqrt{2}g\bar{\mathcal{D}}^2\varphi^a\mathcal{D}([\varphi,\bar{\varphi}]^a) 
	\nonumber \\
	&\quad -\sqrt{2}g\bar{\mathcal{D}}([\varphi,\bar{\varphi}]^a)\mathcal{D}^2\bar{\varphi}^a
	-\sqrt{2}g\bar{\mathcal{D}}([\varphi,\bar{\varphi}]^a)\mathcal{D}\bar{\mathcal{D}}\varphi^a
	+2g^2\bar{\mathcal{D}}([\varphi,\bar{\varphi}]^a)\mathcal{D}([\varphi,\bar{\varphi}]^a) 
	\nonumber \\
	&\quad-\sqrt{2}g\bar{\mathcal{D}}\mathcal{D}\bar{\varphi}^a[\varphi,\mathcal{D}\bar{\varphi}]^a
	-\sqrt{2}g\bar{\mathcal{D}}\mathcal{D}\bar{\varphi}^a[\varphi,\bar{\mathcal{D}}\varphi]^a
	+2g^2\bar{\mathcal{D}}\mathcal{D}\bar{\varphi}^a[\varphi,[\varphi,\bar{\varphi}]]^a 
	\nonumber \\
	&\quad-\sqrt{2}g\bar{\mathcal{D}}^2\varphi^a[\varphi,\mathcal{D}\bar{\varphi}]^a
	-\sqrt{2}g\bar{\mathcal{D}}^2\varphi^a[\varphi,\bar{\mathcal{D}}\varphi]^a
	+2g^2\bar{\mathcal{D}}([\varphi,\bar{\varphi}])[\varphi,\mathcal{D}\bar{\varphi}]^a  
	\nonumber  \\
	&\quad+\sqrt{2}g\mathcal{D}^2\bar{\varphi}^a[\bar{\varphi},\mathcal{D}\bar{\varphi}]^a
	+\sqrt{2}g\mathcal{D}^2\bar{\varphi}^a[\bar{\varphi},\bar{\mathcal{D}}\varphi]^a
	+\sqrt{2}g\mathcal{D}\bar{\mathcal{D}}\varphi^a[\bar{\varphi},\mathcal{D}\bar{\varphi}]^a 
	\nonumber \\
	&\quad+\sqrt{2}g\mathcal{D}\bar{\mathcal{D}}\varphi^a[\bar{\varphi},\bar{\mathcal{D}}\varphi]^a
	-2g^2\mathcal{D}\bar{\mathcal{D}}\varphi^a[\bar{\varphi},[\varphi,\bar{\varphi}]]^a 
	-2g^2\mathcal{D}([\varphi,\bar{\varphi}])[\bar{\varphi},\bar{\mathcal{D}}\varphi]^a  
	\nonumber \\
	&\quad-2g^2[\bar{\varphi},\mathcal{D}\bar{\varphi}]^a[\varphi,\bar{\mathcal{D}}\varphi]^a
	-2g^2[\bar{\varphi},\bar{\mathcal{D}}\varphi]^a[\varphi,\mathcal{D}\bar{\varphi}]^a 
	\nonumber \\
	&\quad+2\sqrt{2}g^2f\bar{\mathcal{D}} ([\varphi,\bar{\varphi}]^a)[\varphi,\delta]^a
	-2\sqrt{2}g^2f\mathcal{D} ([\varphi,\bar{\varphi}]^a)[\bar{\varphi},\delta]^a 
	\nonumber \\
	&\quad-2\sqrt{2}g^2f[\bar{\varphi},\mathcal{D}\bar{\varphi}]^a[\varphi,\delta]^a
	-2\sqrt{2}g^2f[\bar{\varphi},\bar{\mathcal{D}}\varphi]^a[\varphi,\delta]^a 
	\nonumber \\
	&\quad-2\sqrt{2}g^2f[\varphi,\mathcal{D}\bar{\varphi}]^a[\bar{\varphi},\delta]^a
	-2\sqrt{2}g^2f[\varphi,\bar{\mathcal{D}}\varphi]^a[\bar{\varphi},\delta]^a.
	\label{6thterm}
\end{align}
$\delta$ is a Kronecker's delta which appears 
 when $F^a_{56}$ is expanded around the VEV $\langle A_{5, 6} \rangle$ 
 as 
\begin{align}
(\mathcal{D} \bar{\varphi})^a + (\mathcal{\bar{D}} \varphi)^a - \sqrt{2} [\varphi, \bar{\varphi}]^a 
 + \sqrt{2} f \delta^{a1}. 
\end{align}
In these decompositions, 
 we have extracted only the cubic terms with a single $\varphi^1$ or $\bar{\varphi}^1$ 
 and quartic terms with $\varphi^1$ and $\bar{\varphi}^1$, 
 which give contributions to one-loop corrections to the the scalar mass.  
After rewriting the original fields to the fields in the mass eigenstate $\widetilde{A}^a_\mu$, $\widetilde{\varphi}^a$ 
by \eqref{unitaryrot}, \eqref{unitarymatrix} and \eqref{unitaryphi}, 
we expand the terms except for the first term in \eqref{DFDF} in terms of KK modes.  
Using the orthonormality condition for mode functions, 
we obtain four-dimensional interaction terms
\begin{align}
	\mathcal{L}_{\varphi\varphi AA}
	&=8g^2\overline{\varphi}^1\varphi^1\sum_{n,j}\partial_\mu \widetilde{A}^{2}_{\nu,n,j}
	\partial^\mu \widetilde{A}^{2\nu}_{n,j}
	+8g^2\overline{\varphi}^1\varphi^1\sum_{n,j}\partial_\mu \widetilde{A}^{3}_{\nu,n,j}
	\partial^\mu \widetilde{A}^{3\nu}_{n,j} 
	\nonumber \\
	&\quad+16g^2\overline{\varphi}^1\varphi^1\sum_{n,j}\alpha n\widetilde{A}^2_{\mu,n,j} \widetilde{A}^{2\mu}_{n,j}
	+16g^2\overline{\varphi}^1\varphi^1\sum_{n,j}\alpha(n+1)\widetilde{A}^3_{\mu,n,j} \widetilde{A}^{3\mu}_{n,j}, 
	\label{4tenall1} \\
	\mathcal{L}_{\varphi\varphi\varphi\varphi}&=8g^2\overline{\varphi}^1\varphi^1
	\sum_{n,j}\partial_\mu\overline{\widetilde{\varphi}}^2_{n,j}\partial^\mu\widetilde{\varphi}^2_{n,j}
	+8g^2\overline{\varphi}^1\varphi^1\sum_{n,j}\partial_\mu\overline{\widetilde{\varphi}}^3_{n,j}
	\partial^\mu\widetilde{\varphi}^3_{n,j} 
	\nonumber \\
	&\quad+16g^2\overline{\varphi}^1\varphi^1\sum_{n,j}\alpha n\widetilde{\varphi}^2_{n,j}\overline{\widetilde{\varphi}}^2_{n,j}
	+16g^2\overline{\varphi}^1\varphi^1\sum_{n,j}\alpha(n+1)\widetilde{\varphi}^3_{n,j}\overline{\widetilde{\varphi}}^3_{n,j}, 
	\label{4tenall2}
\end{align}
\begin{align}
	\mathcal{L}_{\varphi AA}
	&=+4\sqrt{2}ig\sum_{n,j}\sqrt{\alpha\left(n+1\right)}\partial_\mu\widetilde{A}_{\nu, n, j}^{2} 
	\partial^\mu\widetilde{A}_{n+1,j}^{2 \nu} \overline{\varphi}^{1} 
	-4\sqrt{2}ig\sum_{n, j} \sqrt{\alpha (n+1)} \partial_\mu\widetilde{A}_{\nu, n+1, j}^{3} 
	\partial^\mu\widetilde{A}_{n,j}^{3 \nu} \overline{\varphi}^{1} 
	\nonumber \\
	&\quad -4\sqrt{2}ig\sum_{n, j} \sqrt{\alpha (n+1)} \partial_\mu\widetilde{A}_{\nu, n+1, j}^{2} 
	\partial^\mu\widetilde{A}_{n,j}^{2 \nu} \varphi^{1} 
	+4\sqrt{2}ig\sum_{n, j} \sqrt{\alpha (n+1)} \partial_\mu\widetilde{A}_{\nu, n, j}^{3} 
	\partial^\mu\widetilde{A}_{n+1,j}^{3 \nu} \varphi^{1} 
	\nonumber \\
	&\quad+4\sqrt{2}ig\sum_{n,j}\alpha\left(n+\frac{1}{2}\right)\sqrt{\alpha(n+1)}\widetilde{A}^{2}_{\mu,n+1,j}
	\widetilde{A}^{2\mu}_{n,j}\overline{\varphi}^1 \nonumber \\
	&\quad
	-4\sqrt{2}ig\sum_{n,j}\alpha\left(n+\frac{3}{2}\right)\sqrt{\alpha (n+1)} 
	\widetilde{A}^{3}_{\mu,n,j}\widetilde{A}^{3\mu}_{n+1,j}\overline{\varphi}^1 \nonumber \\
	&\quad-4\sqrt{2}ig\sum_{n,j}\alpha\left(n+\frac{1}{2}\right)\sqrt{\alpha(n+1)}
	\widetilde{A}^{2}_{\mu,n+1,j}\widetilde{A}^{2\mu}_{n,j}\varphi^1 \nonumber \\
	&\quad
	+4\sqrt{2}ig\sum_{n,j}\alpha\left(n+\frac{3}{2}\right)\sqrt{\alpha (n+1)} 
	\widetilde{A}^{3}_{\mu,n,j}\widetilde{A}^{3\mu}_{n+1,j}\varphi^1, 
	\label{3tenall1}
	\end{align}
	\begin{align}
	\mathcal{L}_{\varphi\varphi\varphi}
	&=+4\sqrt{2}ig\sum_{n, j} \sqrt{\alpha (n+1)}\partial_\mu\overline{\widetilde{\varphi}}_{n+1, j}^2 
	\partial^\mu\widetilde{\varphi}_{n, j}^{2} \overline{\varphi}^{1}-4\sqrt{2}ig\sum_{n, j}\sqrt{\alpha  (n+1)}
	\partial_\mu\overline{\widetilde{\varphi}}_{n, j}^3 \partial^\mu\widetilde{\varphi}_{n+1, j}^{3} \overline{\varphi}^{1} 
	\nonumber \\
	&\quad-4\sqrt{2}ig\sum_{n, j}\sqrt{\alpha (n+1)}\partial_\mu\widetilde{\varphi}_{n+1, j}^2 
	\partial^\mu\overline{\widetilde{\varphi}}_{n, j}^{2} \varphi^{1}
	+4\sqrt{2}ig\sum_{n, j} \sqrt{\alpha \left(n+1\right)}\partial_\mu\widetilde{\varphi}_{n, j}^{3} 
	\partial^\mu\overline{\tilde{{\widetilde{\varphi}}}}_{n+1, j}^{3} \varphi^{1} 
	\nonumber \\
	&\quad+4\sqrt{2}ig\sum_{n_2,j}\alpha\left(n-\frac{1}{4}\right)\sqrt{\alpha(n+1)}
	\overline{\widetilde{\varphi}}^2_{n_2+1,j}\widetilde{\varphi}^2_{n_2,j}\overline{\varphi}^1 
	\nonumber \\
	&\quad
	-4\sqrt{2}ig\sum_{n_3,j}\alpha \left(n+\frac{9}{4}\right)\sqrt{\alpha(n+1)}
	\overline{\widetilde{\varphi}}^3_{n_3,j}\widetilde{\varphi}^3_{n_3+1,j}\overline{\varphi}^1 
	\nonumber \\
	&\quad-4\sqrt{2}ig\sum_{n_2,j}\alpha\left(n-\frac{1}{4}\right)\sqrt{\alpha(n+1)}
	\overline{\widetilde{\varphi}}^2_{n_2,j}\widetilde{\varphi}^2_{n_2+1,j}\varphi^1 
	\nonumber \\
	&\quad
	+4\sqrt{2}ig\sum_{n_3,j}\alpha \left(n+\frac{9}{4}\right)\sqrt{\alpha(n+1)}
	\overline{\widetilde{\varphi}}^3_{n_3+1,j}\widetilde{\varphi}^3_{n_3,j}\varphi^1, 
	\label{3tenall2}
\end{align}
where $\alpha$ is $\alpha=2gf$. 
One might think that the finite masses of $\varphi^1, \bar{\varphi}^1$ 
 are generated at one-loop since all of the interactions \eqref{4tenall1}
 --\eqref{3tenall2} 
 have the nonderivative terms for $\varphi^1, \bar{\varphi}^1$. 
However, these nonderivative terms are originated from the commutators 
 such as $[\quad, \varphi^1], [\quad, \bar{\varphi}^1]$ and $[\varphi^1, \bar{\varphi}^1]$. 
It means that these are invariant under the constant shifts for $\varphi^1, \bar{\varphi}^1$, 
 which are the transformations of translation in compactified space. 
Therefore, the one-loop corrections to the scalar masses are still expected to be vanished. 

Four-dimensional interaction Lagrangian is summarized as
\begin{align}
	\mathcal{L}_{4,int}=\frac{1}{\Lambda^2}(\mathcal{L}_{\varphi\varphi AA}
	+\mathcal{L}_{\varphi\varphi\varphi\varphi}
	+\mathcal{L}_{\varphi AA}+\mathcal{L}_{\varphi\varphi\varphi}).
	\label{lagall}
\end{align}

\section{One-loop Corrections to Scalar Masses from Higher Dimensional Operators}
In this section, 
 we calculate one-loop corrections to the zero mode of the scalar fields $\varphi^1, \bar{\varphi}^1$, 
which arise from the extra components of the gauge fields $A_{5,6}$. 
In our theory, 
 those corrections can be calculated from two kinds of the diagrams. 
One is one-loop diagrams from a quartic interaction term, 
 the other is those from two cubic interaction terms. 
As was discussed in \cite{HM}, 
 the zero mode of the scalar fields $\varphi^1, \bar{\varphi}^1$ can be regarded as 
 Nambu-Goldstone (NG) bosons with respect to the spontaneous breaking 
 of translational symmetry in compactified space. 
Therefore, $\varphi^1, \bar{\varphi}^1$ are allowed to have only derivative interactions
 and expected to be massless. 
In \cite{HM}, we have shown that this expectation is indeed correct 
 by computing one-loop corrections to the masses of $\varphi^1, \bar{\varphi}^1$ explicitly 
 in a six-dimensional Yang-Mills theory. 
Although the only renormalizable terms were considered in our previous paper \cite{HM}, 
 we have to take into account all of the higher dimensional interactions 
 consistent with symmetry of the theory since our theory is non-renormalizable. 

We will explicitly show below that one-loop corrections to the scalar masses are indeed cancelled
 even if the lowest term of the higher dimensional operators are present. 
The statement is straightforward, 
 but the cancellation of quantum corrections to scalar mass is somewhat nontrivial 
 since the cancelation is realized among the terms with different orders of $1/\Lambda^2$. 
For detail calculations, see Appendices. 

\subsection{One-loop Corrections from the Quartic Interactions}
\begin{figure}[h]
\begin{center}
\includegraphics[width=3.5cm, bb=9 9 163 181]{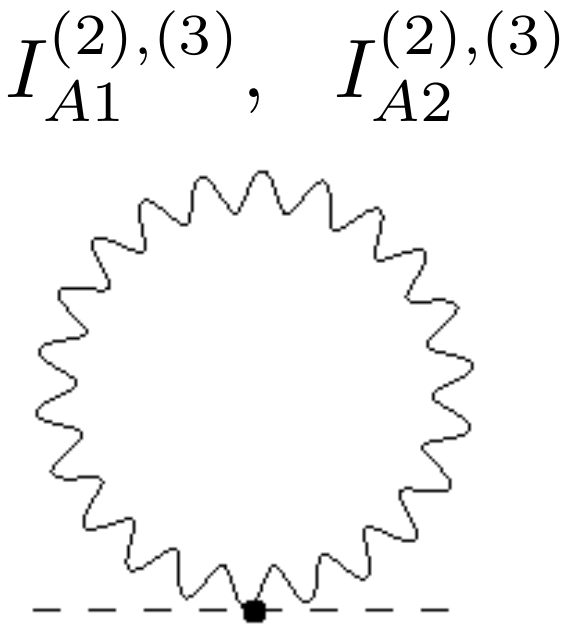}
\caption{One-loop diagram from the gauge boson loops by the quartic interaction.}
\label{fig1}
\end{center}
\end{figure}
From the interactions in \eqref{4tenall1}, 
there are four types of one-loop corrections to the scalar masses 
 from the gauge boson loop contributions as in Figure \ref{fig1}, 
 which are expressed as
\begin{align}
	I^{(2)}_{A1}&=\frac{32ig^2}{\Lambda^2} |N| \sum_{n=0}^\infty\int\frac{d^4p}{(2\pi)^4}\frac{p^2}{p^2+\alpha n}, 
	\label{A1(2)} \\
	I^{(3)}_{A1}&= \frac{32ig^2}{\Lambda^2} |N|\sum_{n=0}^\infty\int\frac{d^4p}{(2\pi)^4}\frac{p^2}{p^2+\alpha (n+1)}, 
	\label{A1(3)} \\
	I^{(2)}_{A2}&= \frac{64ig^2}{\Lambda^2} |N|\sum_{n=0}^\infty\int\frac{d^4p}{(2\pi)^4}\frac{\alpha n}{p^2+\alpha n}, 
	\label{A2(2)} \\
	I^{(3)}_{A2}&= \frac{64ig^2}{\Lambda^2}|N|\sum_{n=0}^\infty\int\frac{d^4p}{(2\pi)^4}\frac{\alpha (n+1)}{p^2+\alpha(n+1)}, 
	\label{A2(3)}
\end{align}
where Wick rotation in momentum integral is understood throughout this paper. 
The superscript (2), (3) means the contributions from $\widetilde{A}^2_{\mu},\widetilde{A}^3_{\mu}$ loops respectively. 
$A1$ and $A2$ represent corrections from the interactions 
 in the first line and the second line of \eqref{4tenall1}, respectively. 
Performing the dimensional regularization for the four dimensional momentum integral, 
 we find 
\begin{align}
	I_{A1}&\equiv I^{(2)}_{A1}+ I^{(3)}_{A1}
	=-i\frac{4g^2\alpha^2|N|}{\pi^2 \Lambda^2}\left(\frac{4\pi}{\alpha}\right)^{\epsilon}\Gamma(\epsilon-1)\zeta[\epsilon-2,0], 
	\label{IA1} \\
	I_{A2}&\equiv I^{(2)}_{A2}+I^{(3)}_{A2} 
	=+i\frac{8g^2\alpha^2|N|}{\pi^2 \Lambda^2}\left(\frac{4\pi}{\alpha}\right)^\epsilon\Gamma(\epsilon-1)\zeta[\epsilon-2,0], 
	\label{IA2}
\end{align}
where $\zeta[s,a]$ is Hurwitz zeta function and 
 $\epsilon$ is defined in the ordinary dimensional regularization as $d=4-2\epsilon$. 

Summing up \eqref{IA1} and \eqref{IA2}, 
 we obtain the total gauge boson loop contributions to one-loop correction due to the quartic interactions.  
\begin{align}
	I_{\varphi\varphi AA}\equiv I_{A1}+I_{A2}
	=+i\frac{4g^2\alpha^2|N|}{\pi^2 \Lambda^2}\left(\frac{4\pi}{\alpha}\right)^{\epsilon}\Gamma(\epsilon-1)\zeta[\epsilon-2,0].
	\label{result4A}
\end{align}
\begin{figure}[h]
\begin{center}
\includegraphics[width=3cm, bb=9 9 163 184]{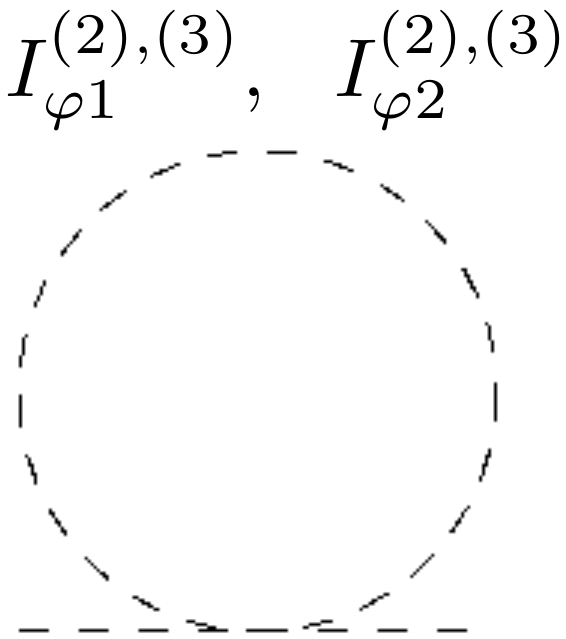}
\caption{One-loop diagram from the scalar loops by the quartic interaction.}
\label{fig2}
\end{center}
\end{figure}

Next, we consider the corrections from the scalar quartic interactions \eqref{4tenall2}.
There are also four types of one-loop corrections to the scalar masses 
 from the scalar loop contributions as in Figure \ref{fig2}, which are expressed as
\begin{align}
	I^{(2)}_{\varphi 1}&= \frac{8ig^2}{\Lambda^2} |N|\sum_{n=0}^\infty\int\frac{d^4p}{(2\pi)^4}
	\frac{p^2}{p^2+\alpha\left(n+\frac{1}{2}\right)}, 
	\label{phi1(2)} \\
	I^{(3)}_{\varphi1}&= \frac{8ig^2}{\Lambda^2} |N|\sum_{n=0}^\infty\int\frac{d^4p}{(2\pi)^4}
	\frac{p^2}{p^2+\alpha\left(n+\frac{1}{2}\right)}, 
	\label{phi1(3)} \\
	I^{(2)}_{\varphi2}&= \frac{16ig^2}{\Lambda^2}|N|\sum_{n=0}^\infty\int\frac{d^4p}{(2\pi)^4}
	\frac{\alpha n}{p^2+\alpha\left(n+\frac{1}{2}\right)}, 
	\label{phi2(2)} \\
	I^{(3)}_{\varphi2}&= \frac{16ig^2}{\Lambda^2}|N|\sum_{n=0}^\infty\int\frac{d^4p}{(2\pi)^4}
	\frac{\alpha(n+1)}{p^2+\alpha\left(n+\frac{1}{2}\right)}, 
	\label{phi2(3)}
\end{align}
where the superscript (2), (3) means the contributions from 
 $\widetilde{\varphi}^2,\widetilde{\varphi}^3$ loops respectively.  
$\varphi1$ and $\varphi2$ represent corrections from the interactions 
 in the first line and the second line of \eqref{4tenall2}, respectively. 
Calculating these corrections similarly to the above gauge boson loop, 
we obtain 
\begin{align}
	I_{\varphi1}&\equiv I^{(2)}_{\varphi1}+ I^{(3)}_{\varphi1}
	=-i\frac{g^2\alpha^2|N|}{\pi^2 \Lambda^2}\left(\frac{4\pi}{\alpha}\right)^\epsilon\Gamma(\epsilon-1)\zeta[\epsilon-2,1/2], 
	\label{Iphi1}\\
	I_{\varphi2}&\equiv I^{(2)}_{\varphi2}+I^{(3)}_{\varphi2}
	=+i\frac{2g^2\alpha^2|N|}{\pi^2 \Lambda^2}\left(\frac{4\pi}{\alpha}\right)^\epsilon\Gamma(\epsilon-1)\zeta[\epsilon-2,1/2]. 
	\label{Iphi2}
\end{align}
Summing up \eqref{Iphi1} and \eqref{Iphi2}, 
we obtain the total scalar loop contributions to one-loop correction due to the scalar quartic interactions. 
\begin{align}
	I_{\varphi\varphi\varphi\varphi}\equiv I_{\varphi1}+I_{\varphi2}
	=+i\frac{g^2\alpha^2|N|}{\pi^2 \Lambda^2}\left(\frac{4\pi}{\alpha}\right)^\epsilon\Gamma(\epsilon-1)\zeta[\epsilon-2,1/2].
	\label{result4p}
\end{align}

\subsection{One-loop Corrections from the Cubic Interactions} 
In the case of the corrections due to the cubic interactions, 
 we note that one-loop corrections 
 using both the cubic interaction in $\mathcal{O}(\Lambda^0)$ and that in $\mathcal{O}(1/\Lambda^{2})$ appear. 
This is a nontrivial point in calculating the corrections in the presence of the higher dimensional operators.  
\begin{figure}[h]
\begin{center}
\includegraphics[width=6cm, bb=9 9 271 183]{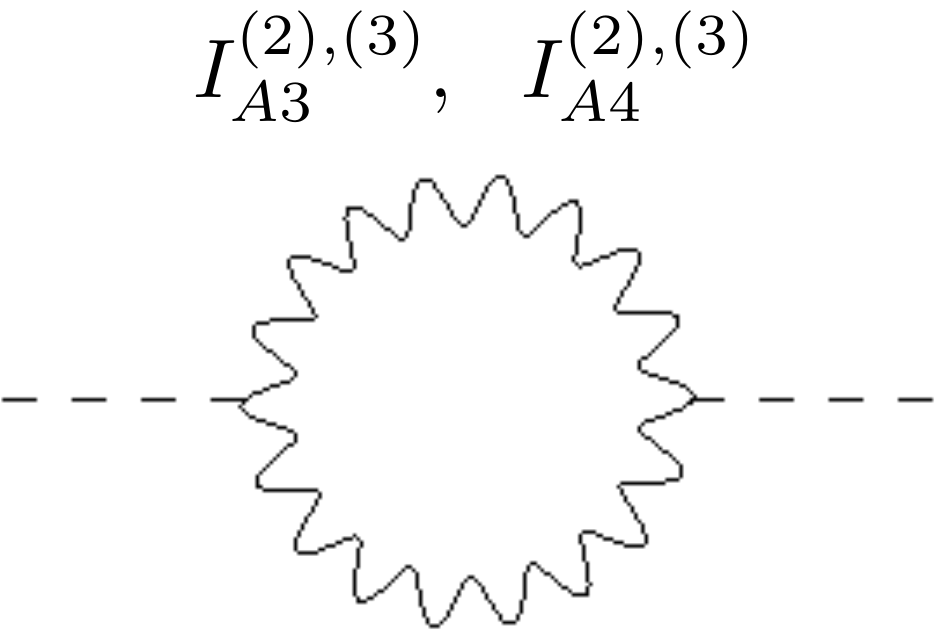}
\caption{One-loop diagram from the gauge boson loops by the cubic interactions.}
\label{fig3}
\end{center}
\end{figure}
From the interacitons \eqref{paperphiAA} and \eqref{3tenall1}, 
 there are four types of one-loop corrections to the scalar masses 
 from the gauge boson loop contributions as in Figure \ref{fig3} . 
\begin{align}
	I^{(2)}_{A3}&=- \frac{32ig^2}{\Lambda^2} |N|\sum_{n=0}^\infty\int \frac{d^{4} p}{(2 \pi)^{4}}
	\frac{\alpha(n+1)p^2}{(p^2+\alpha n)(p^2+\alpha(n+1))}, 
	\label{A3(2)} \\
	I^{(3)}_{A3}&=- \frac{32ig^2}{\Lambda^2}|N|\sum_{n=0}^\infty\int \frac{d^{4} p}{(2 \pi)^{4}}
	\frac{\alpha(n+1)p^2}{(p^2+\alpha (n+1))(p^2+\alpha(n+2))}, 
	\label{A3(3)} \\
	I^{(2)}_{A4}&=-\frac{32ig^2}{\Lambda^2} |N|\sum_{n=0}^\infty\int\frac{d^4p}{(2\pi)^4}
	\frac{\alpha(n+1) \alpha\left(n+\frac{1}{2}\right)}{(p^2+\alpha n)(p^2+\alpha(n+1))}, 
	\label{A4(2)} \\
	I^{(3)}_{A4}&=-\frac{32ig^2}{\Lambda^2}|N|\sum_{n=0}^\infty\int\frac{d^4p}{(2\pi)^4}
	\frac{\alpha(n+1) \alpha\left(n+\frac{3}{2}\right)}{(p^2+\alpha (n+1))(p^2+\alpha(n+2))}. 
	\label{A4(3)}
\end{align}
where $A3 (A4)$ represents the contributions from the interactions of 
 the first and the second (from the third to the sixth) lines in \eqref{3tenall1}, respectively.  
Calculating these corrections by dimensional regularization, 
we find 
\begin{align}
	I_{A3}&\equiv I^{(2)}_{A3}+ I^{(3)}_{A3}
	=+i\frac{4g^2\alpha^2|N|}{\pi^2 \Lambda^2}\left(\frac{4\pi}{\alpha}\right)^\epsilon 
	\Gamma(\epsilon-1) \zeta[\epsilon-2,0], 
	\label{IA3} \\
	I_{A4}&\equiv I^{(2)}_{A4}+I^{(3)}_{A4}
	=-i\frac{8g^2\alpha^2|N|}{\pi^2 \Lambda^2}\left(\frac{4\pi}{\alpha}\right)^\epsilon
	\Gamma(\epsilon-1)\zeta[\epsilon-2,0]. 
	\label{IA4}
\end{align}
Summing up these results \eqref{IA3} and \eqref{IA4}, 
 we obtain the total one-loop corrections to the scalar masses from the gauge boson loop contributions. 
	\begin{align}
	I_{\varphi AA}\equiv I_{A3}+I_{A4} 
	=-i\frac{4g^2\alpha^2|N|}{\pi^2 \Lambda^2}\left(\frac{4\pi}{\alpha}\right)^\epsilon\Gamma(\epsilon-1)\zeta[\epsilon-2,0].
	\label{result3A}
	\end{align}

\begin{figure}[h]
\begin{center}
\includegraphics[width=5cm, bb=9 9 262 184]{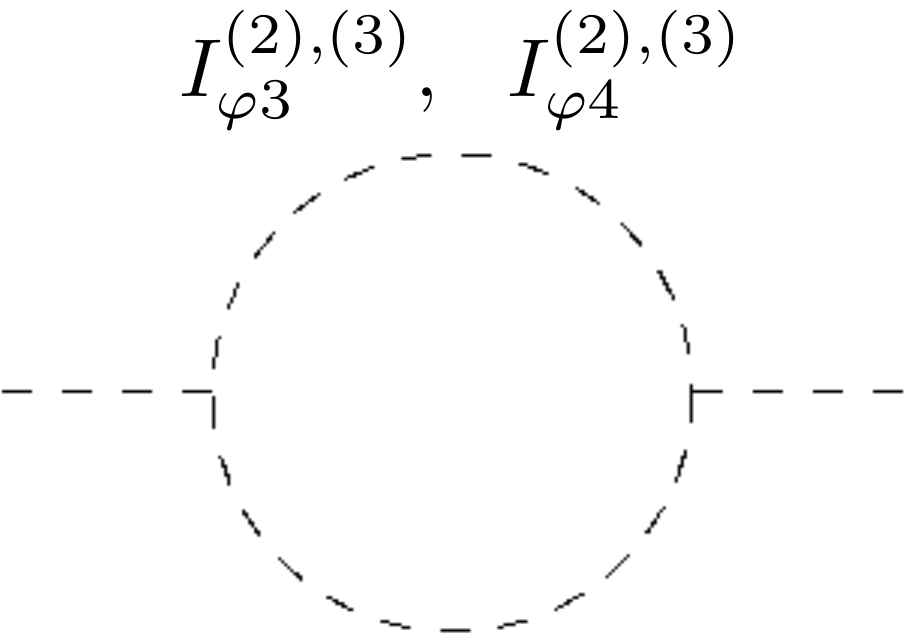}
\caption{One-loop diagram from the scalar loops by the cubic interactions.}
\label{fig4}
\end{center}
\end{figure}
Next, we consider the corrections from \eqref{paperphiphi} and \eqref{3tenall2}, 
 which also give four types of one-loop corrections 
 from the scalar loop contributions as in Figure \ref{fig4}. 
	\begin{align}
	I^{(2)}_{\varphi3}&=-\frac{8ig^{2}}{\Lambda^2}|N| \sum_{n=0}^{\infty} \int \frac{d^{4} p}{(2 \pi)^{4}} 
	\frac{\alpha(n+1) p^{2}}{\left(p^{2}+\alpha\left( n+\frac{1}{2}\right)\right)\left(p^{2}
	+\alpha\left(n+\frac{3}{2}\right)\right)}, 
	\label{phi3(2)} \\
	I^{(3)}_{\varphi3}&=- \frac{8i g^{2}}{\Lambda^2}|N| \sum_{n=0}^{\infty} \int \frac{d^{4} p}{(2 \pi)^{4}} 
	\frac{\alpha(n+1) p^{2}}{\left(p^{2}+\alpha\left( n+\frac{1}{2}\right)\right)
	\left(p^{2}+\alpha\left(n+\frac{3}{2}\right)\right)}, 
	\label{phi3(3)} \\
	I^{(2)}_{\varphi4}&=- \frac{8i g^{2}}{\Lambda^2}|N| \sum_{n=0}^{\infty} \int \frac{d^{4} p}{(2 \pi)^{4}} 
	\frac{\alpha(n+1)\alpha\left(n-\frac{1}{4}\right)}{\left(p^{2}+\alpha\left( n+\frac{1}{2}\right)\right)
	\left(p^{2}+\alpha\left(n+\frac{3}{2}\right)\right)}, 
	\label{phi4(2)} \\
	I^{(3)}_{\varphi4}&=-\frac{8i g^{2}}{\Lambda^2}|N| \sum_{n=0}^{\infty} \int \frac{d^{4} p}{(2 \pi)^{4}} 
	\frac{\alpha(n+1)\alpha\left(n+\frac{9}{4}\right)}{\left(p^{2}+\alpha\left( n+\frac{1}{2}\right)\right)
	\left(p^{2}+\alpha\left(n+\frac{3}{2}\right)\right)}.  
	\label{phi4(3)}
	\end{align}
$\varphi3 (\varphi4)$ represents the contributions from the interactions of 
 the first and the second (from the third to the sixth) lines in \eqref{3tenall2}, respectively.  
By similar calculations, we find 
\begin{align}
	I_{\varphi3}&\equiv I^{(2)}_{\varphi3}+I^{(3)}_{\varphi3}
	=+i\frac{g^2\alpha^2|N|}{\pi^2 \Lambda^2}\left(\frac{4\pi}{\epsilon}\right)^\epsilon\Gamma(\epsilon-1)\zeta[\epsilon-2,1/2], 
	\label{Iphi3}\\
	I_{\varphi4}&\equiv I^{(2)}_{\varphi4}+I^{(3)}_{\varphi4}
	=-i\frac{2g^2\alpha^2|N|}{\pi^2 \Lambda^2}\left(\frac{4\pi}{\alpha}\right)^\epsilon\Gamma(\epsilon-1)\zeta[\epsilon-2,1/2]. 
	\label{Iphi4}
\end{align}
Summing up \eqref{Iphi3} and \eqref{Iphi4}, 
 we obtain the total one-loop corrections to the scalar masses from the scalar loop contributions. 
\begin{align}
	I_{\varphi\varphi\varphi}&\equiv I_{\varphi3}+I_{\varphi4}
	=-i\frac{g^2\alpha^2|N|}{\pi^2 \Lambda^2}\left(\frac{4\pi}{\alpha}\right)^\epsilon\Gamma(\epsilon-1)\zeta[\epsilon-2,1/2]. 
	\label{result3p}
\end{align}

\subsection{Cancellation of One-loop Corrections to Scalar Mass at $\mathcal{O}(1/\Lambda^{2})$}
Summing up all of the results \eqref{result4A}, \eqref{result4p}, \eqref{result3A} and \eqref{result3p}, 
 we can verify that one-loop corrections to the scalar masses are indeed cancelled at the leading order of $1/\Lambda^2$. 
\begin{align}
	&I_{\varphi \varphi AA}+I_{\varphi AA}=0, \\
	&I_{\varphi\varphi\varphi\varphi}+I_{\varphi\varphi\varphi}=0.
\end{align}
As can be seen, 
 the gauge loop contributions and the scalar loop contributions are independently cancelled. 
In particular, the scalar loop contributions can be cancelled without the ghost loop contributions, 
 which is different from the case of Yang-Mills theory \cite{HM}.

\subsection{Comments on the Corrections from the Higher Dimensional Operators 
More Than $\mathcal{O}(1/\Lambda^{4})$}
Finally, we would like to comment on the generalization of our discussion, 
 namely, the corrections from the higher dimensional operators more than $\mathcal{O}(1/\Lambda^{4})$. 
In this case, the cancellation of the one-loop corrections to the scalar masses becomes more involved. 
For instance,  let us consider the next to leading order higher dimensional operators \eqref{O2}. 
We immediately find that the first term vanishes 
 because of the traceless condition for SU(2) generators as in the section \ref{DDDF} 
 and the third term also vanishes because of the properties of totally anti-symmetric tensor 
 and the trace of generators as in the section \ref{FFF}.
Thus, $\mathcal{O}_2(D,F)$ is reduced to 
	\begin{align}
	\mathcal{O}_2(D,F)&=\mathrm{Tr}[D_kD_LF_{MN}D^KD^LF^{MN}]+\mathrm{Tr}[F_{MN}F^{MN}F_{AB}F^{AB}].
	\label{O2ver2}
	\end{align}

If we use two kinds of cubic interactions \eqref{3tenall1} and \eqref{3tenall2}, 
 we obtain some one-loop corrections at the second order of $1/\Lambda^2$, 
 however these corrections are not cancelled 
 because we must take into account the contributions 
 with the operators of $\mathcal{O}(1/\Lambda^0)$ and $\mathcal{O}(1/\Lambda^{4})$.
Even at $\mathcal{O}(1/\Lambda^{4})$, 
 although it is relatively easy to calculate the second term in \eqref{O2ver2}, 
 the first term in \eqref{O2ver2} is found to have huge number of interaction terms 
 which are relevant to the one-loop corrections to the scalar masses. 
At higher order than $\mathcal{O}(1/\Lambda^{4})$, 
 we need to consider carefully the variety of combinations 
 among the operators different order of $1/\Lambda^2$ and it becomes more complicated.  
Such an analysis is very interesting, 
 however it is beyond the scope of this paper and we leave it for a future study.

\section{Summary}
Since the higher dimensional theory is nonrenormalizable, 
 the higher dimensional operators consistent with symmetry of the theory 
 should be taken into account in the Lagrangian. 
In this paper, we have shown that one-loop corrections to the masses of the scalar fields, 
 which are zero modes of extra components of the higher dimensional gauge fields, 
 are indeed cancelled in flux compactification of 
 a six dimensional SU(2) Yang-Mills theory including the leading order of higher dimensional operators. 
Even if the higher dimensional operators are taken into account, 
 the cancellation is expected to be true as long as the zero mode of the scalar fields are NG bosons 
 with respect to the translational symmetry in compactified space. 
The statement is straightforward, 
 but the cancellation itself is not so trivial since the contributions to the scalar masses 
 come from the interactions in the different order of the higher dimensional operators. 
In fact, the one-loop corrections to the scalar masses at ${\cal O}(1/\Lambda^2)$ are generated 
 from the cubic interactions of higher dimensional operators at ${\cal O}(1/\Lambda^0)$ and ${\cal O}(1/\Lambda^2)$ 
 and were shown to be cancelled in this paper. 
This cancellation would not be changed 
 even if we take into account the fermion contributions.  

One of the interesting applications of our results obtained in this paper is
 to identify the scalar field with the SM Higgs field. 
Since the information on the new physics are encoded 
 in the higher dimensional operators consistent with the SM gauge symmetry, 
 our observations and results in this paper are quite useful. 
As mentioned in the introduction, 
 since the Higgs field is massless as it stands, 
 the translational symmetry in compactified space must be explicitly broken 
 at weak scale to give a mass to the Higgs field. 
Understanding a mechanism to realize this situation is a next issue in our future study.

\section*{Acknowledgments}
This work is supported in part by JSPS KAKENHI Grant Number JP17K05420 (N.M.).

\appendix
\section{Detail calculations for one-loop corrections to the scalar masses}
In this Appendix, the calculations in the main text are explained in detail \cite{Lee, BHT}. 
 
\subsection{Loop Integral Fomula and Hurwitz Zeta Function}\label{LFZF}
In the calculation of the loop integral, 
 we employ the dimensional regularization and use these integrals in Euclidean space.
\begin{align}
	\int \frac{d^{d} p}{(2 \pi)^{d}} \frac{1}{\left(p^{2}+\Delta\right)^{n}}
	&=\frac{1}{(4 \pi)^{d / 2}} \frac{\Gamma\left(n-\frac{d}{2}\right)}{\Gamma(n)}\left(\frac{1}{\Delta}\right)^{n-\frac{d}{2}}, 
	\label{DR1}\\
	\int \frac{d^{d} p}{(2 \pi)^{d}} \frac{p_{\mu} p_{\nu}}{\left(p^{2}+\Delta\right)^{n}}
	&=\frac{1}{(4 \pi)^{d / 2}} \frac{\delta_{\mu \nu}}{2} \frac{\Gamma\left(n-\frac{d}{2}-1\right)}{\Gamma(n)}
	\left(\frac{1}{\Delta}\right)^{n-\frac{d}{2}-1},
	\label{DR2}
\end{align}
where $d=4-2\epsilon$ and $\Gamma(n)$ is gamma function.

As for the mode summation with respect to $n$, 
 it is convenient to use Hurwitz zeta function
	\begin{align}
	\zeta[s,a]=\sum_{n=0}^\infty\frac{1}{(n+a)^s}.
	\end{align}
In the main text, we deal with $\zeta[\epsilon-2,a]$ and $\zeta[\epsilon-3,a]$ $(a=0,1/2,1,3/2,2)$, 
therefore, we summarize these values.
\begin{align*}
	\zeta[\epsilon-2,0]&=\frac{\zeta(3)}{4\pi^2}\epsilon,
	&\zeta[\epsilon-3,0]&=\frac{1}{120}+\zeta'(-3)\epsilon \\
	\zeta[\epsilon-2,1/2]&=\frac{3\zeta(3)}{16\pi^2}\epsilon,
	&\zeta[\epsilon-3,1/2]&=-\frac{7}{960}+\left(\frac{\ln2}{960}-\frac{7}{8}\zeta'(-3)\right)\epsilon \\
	\zeta[\epsilon-2,1]&=\frac{\zeta(3)}{4\pi^2}\epsilon,
	&\zeta[\epsilon-3,1]&=\frac{1}{120}+\zeta'(-3)\epsilon \\
	\zeta[\epsilon-2,3/2]&=-\frac{1}{4}+\left(-\frac{\ln2}{4}+\frac{3\zeta(3)}{16\pi^2}\right)\epsilon,
	&\zeta[\epsilon-3,3/2]&=-\frac{127}{960}+\left(-\frac{119\ln2}{960}-\frac{7}{8}\zeta'(-3)\right)\epsilon \\
	\zeta[\epsilon-2,2]&=-1+\frac{\zeta(3)}{4\pi^2}\epsilon,
	&\zeta[\epsilon-3,2]&=-\frac{119}{120}+\zeta'(-3)\epsilon
\end{align*}

\subsection{Case \eqref{A1(2)} and \eqref{A1(3)}} \label{appA1}
To calculate \eqref{A1(2)}, $J$ is defined as
\begin{align}
	J\equiv\sum_{n=0}^\infty\int\frac{d^4p}{(2\pi)^4}\frac{p^2}{p^2+\alpha n}.
\end{align}
Using \eqref{DR2}, $J$ can be expressed as
\begin{align}
	J&=\sum_{n=0}^\infty\frac{1}{(4 \pi)^{d / 2}} \frac{d}{2} 
	\Gamma\left(-\frac{d}{2}\right)\left(\frac{1}{\alpha n}\right)^{-\frac{d}{2}} \nonumber \\
	&=-\frac{\alpha^2}{16\pi^2}\left(\frac{4\pi}{\alpha}\right)^{\epsilon}\Gamma(\epsilon-1)\zeta[\epsilon-2,0].
\end{align}
Similarly, to calculate \eqref{A1(3)}, $J'$ is defined as and using \eqref{DR2}, we find
\begin{align}
	J'&\equiv\sum_{n=0}^\infty\int\frac{d^4p}{(2\pi)^4}\frac{p^2}{p^2+\alpha (n+1)} \nonumber \\
	&=-\frac{\alpha^2}{16\pi^2}\left(\frac{4\pi}{\alpha}\right)^{\epsilon}\Gamma(\epsilon-1)\zeta[\epsilon-2,1].
\end{align}
Noting the fact that $\zeta[\epsilon-2,1]=\zeta[\epsilon-2,0]$, 
 $J$ and $J'$ are found to be same.
Thus, $I_{A1}$ is computed as
\begin{align}
	I_{A1}&\equiv I^{(2)}_{A1}+ I^{(3)}_{A1}=\frac{64ig^2}{\Lambda^2}|N|J \nonumber \\
	&=-i\frac{4g^2\alpha^2|N|}{\pi^2 \Lambda^2}\left(\frac{4\pi}{\alpha}\right)^{\epsilon}\Gamma(\epsilon-1)\zeta[\epsilon-2,0]. 
\end{align}

\subsection{Case \eqref{A2(2)} and \eqref{A2(3)}} \label{appA2}
To calculate \eqref{A2(2)}, $K$ is defined as
\begin{align}
	K\equiv\sum_{n=0}^\infty\alpha n\int\frac{d^4p}{(2\pi)^4}\frac{1}{p^2+\alpha n}.
\end{align}
Using \eqref{DR1}, $K$ can be expressed as 
\begin{align}
	K&=\sum_{n=0}^\infty\alpha n\frac{1}{(4\pi)^{d/2}}
	\Gamma\left(1-\frac{d}{2}\right)\left(\frac{1}{\alpha n}\right)^{1-d/2} \nonumber \\
	&=\frac{\alpha^2}{(4\pi)^2}\left(\frac{4\pi}{\alpha}\right)^\epsilon\Gamma(\epsilon-1)\zeta[\epsilon-2,0].
\end{align}
Similarly, to calculate \eqref{A2(3)}, $K'$ is defined as and using \eqref{DR1}, we find
\begin{align}
	K'&\equiv\sum_{n=0}^\infty\alpha (n+1)\int\frac{d^4p}{(2\pi)^4}\frac{1}{p^2+\alpha (n+1)}
	 \nonumber \\
	&=\frac{\alpha^2}{(4\pi)^2}\left(\frac{4\pi}{\alpha}\right)^\epsilon\Gamma(\epsilon-1)\zeta[\epsilon-2,1].
\end{align}
Because of $\zeta[\epsilon-2,1]=\zeta[\epsilon-2,0]$ again, $K$ and $K'$ are found to be same.
Thus, $I_{A2}$ is computed as
\begin{align}
	I_{A2}&\equiv I^{(2)}_{A2}+I^{(3)}_{A2}=\frac{128ig^2}{\Lambda^2}|N|K \nonumber \\
	&=+i\frac{8g^2\alpha^2|N|}{\pi^2 \Lambda^2}\left(\frac{4\pi}{\alpha}\right)^\epsilon\Gamma(\epsilon-1)\zeta[\epsilon-2,0]. 
\end{align}

\subsection{Case \eqref{phi1(2)} and \eqref{phi1(3)}} \label{appphi1}
To calculate \eqref{phi1(2)} and \eqref{phi1(3)}, $L$ is defined as
\begin{align}
	L\equiv\sum_{n=0}^\infty\int\frac{d^4p}{(2\pi)^4}\frac{p^2}{p^2+\alpha \left(n+\frac{1}{2}\right)}.
\end{align}
Using \eqref{DR2}, $L$ is expressed to be 
\begin{align}
	L&=\sum_{n=0}^\infty\frac{1}{(4 \pi)^{d / 2}} \frac{d}{2} 
	\Gamma\left(-\frac{d}{2}\right)\left(\frac{1}{\alpha \left(n+\frac{1}{2}\right)}\right)^{-\frac{d}{2}} \nonumber \\
	&=-\frac{\alpha^2}{16\pi^2}\left(\frac{4\pi}{\alpha}\right)^{\epsilon}\Gamma(\epsilon-1)\zeta[\epsilon-2,1/2].
\end{align}
Thus, $I_{\varphi1}$ is computed as
\begin{align}
	I_{\varphi1}&\equiv I^{(2)}_{\varphi1}+ I^{(3)}_{\varphi1}=\frac{16ig^2}{\Lambda^2}|N|L \nonumber \\
	&=-i\frac{g^2\alpha^2|N|}{\pi^2 \Lambda^2}\left(\frac{4\pi}{\alpha}\right)^\epsilon
	\Gamma(\epsilon-1)\zeta[\epsilon-2,1/2]. 
\end{align}

\subsection{Case \eqref{phi2(2)} and \eqref{phi2(3)}} \label{appphi2}
In this calculation, 
 it is easy to calculate $I_{\varphi2}$ defined as $I_{\varphi2}\equiv I^{(2)}_{\varphi2}+I^{(3)}_{\varphi2}$ 
 rather than computations of $I^{(2)}_{\varphi2}$ and $I^{(3)}_{\varphi2}$ individually.
Then, $I_{\varphi2}$ is given by
\begin{align}
	I_{\varphi2}&=I^{(2)}_{\varphi2}+I^{(3)}_{\varphi2} \nonumber \\
	&=\frac{32ig^2}{\Lambda^2}|N|\sum_{n=0}^\infty\int\frac{d^4p}{(2\pi)^4}
	\frac{\alpha\left(n+\frac{1}{2}\right)}{p^2+\alpha\left(n+\frac{1}{2}\right)}.
\end{align}
To calculate $I_{\varphi2}$, $M$ is defined as
\begin{align}
	M\equiv\sum_{n=0}^\infty\int\frac{d^4p}{(2\pi)^4}
	\frac{\alpha\left(n+\frac{1}{2}\right)}{p^2+\alpha\left(n+\frac{1}{2}\right)}.
\end{align}
Using \eqref{DR1}, $M$ expresses
\begin{align}
	M=\frac{\alpha^2}{16\pi^2}\left(\frac{4\pi}{\alpha}\right)^\epsilon\Gamma(\epsilon-1)\zeta[\epsilon-2,1/2]
\end{align}
Thus, $I_{\varphi2}$ is computed as
\begin{align}
	I_{\varphi2}&=\frac{32ig^2}{\Lambda^2}|N|M \nonumber \\
	&=+i\frac{2g^2\alpha^2|N|}{\pi^2 \Lambda^2}\left(\frac{4\pi}{\alpha}\right)^\epsilon\Gamma(\epsilon-1)\zeta[\epsilon-2,1/2]. 
\end{align}

\subsection{Case \eqref{A3(2)} and \eqref{A3(3)}} \label{appA3}
To calculate \eqref{A3(2)}, $A$ is defined as
\begin{align}
	A&\equiv\sum_{n=0}^\infty\int \frac{d^{4} p}{(2 \pi)^{4}}\frac{\alpha(n+1)p^2}{(p^2+\alpha n)(p^2+\alpha(n+1))} \nonumber \\
	&=\sum_{n=0}^\infty\alpha(n+1)\int_0^1dx\int \frac{d^{4} p}{(2 \pi)^{4}}\frac{p^2}{(p^2+\Delta(x,n))^2},
\end{align}
where $\Delta(x,n)=\alpha(n+1)-\alpha x$.
Using \eqref{DR2}, $A$ is expressed as 
\begin{align}
	A&=\sum_{n=0}^\infty\alpha(n+1)\frac{1}{(4 \pi)^{d / 2}} 
	\frac{d}{2} \Gamma\left(1-\frac{d}{2}\right)\left(\frac{1}{\Delta}\right)^{1-\frac{d}{2}} 
	\nonumber \\
	&=(-1)^{\epsilon-1}\frac{\alpha^2}{16\pi^2}\left(\frac{4\pi}{\alpha}\right)^\epsilon(2-\epsilon)
	\Gamma(\epsilon-1)\sum_{n=0}^\infty(n+1)\int_0^1dx\left(\frac{1}{x-\alpha(n+1)}\right)^{\epsilon-1}.
\end{align}
Performing the integral over $x$, we find 
\begin{align}
	A&=-\frac{\alpha^2}{16\pi^2}\left(\frac{4\pi}{\alpha}\right)^\epsilon\Gamma(\epsilon-1)
	\sum_{n=0}^\infty(n+1)\left(\frac{1}{n^{\epsilon-2}}-\frac{1}{(n+1)^{\epsilon-2}}\right) \nonumber \\
	&=-\frac{\alpha^2}{16\pi^2}\left(\frac{4\pi}{\alpha}\right)^\epsilon\Gamma(\epsilon-1)
	\Big(\zeta[\epsilon-3,0]+\zeta[\epsilon-2,0]-\zeta[\epsilon-3,1]\Big) \nonumber \\
	&=-\frac{\alpha^2}{16\pi^2}\left(\frac{4\pi}{\alpha}\right)^\epsilon \Gamma(\epsilon-1)	\zeta[\epsilon-2,0].
\end{align}
Similarly, to calculate \eqref{A3(3)}, $A'$ is defined as
\begin{align}
	A'&\equiv\sum_{n=0}^\infty\int \frac{d^{4} p}{(2 \pi)^{4}}
	\frac{\alpha(n+1)p^2}{(p^2+\alpha (n+1))(p^2+\alpha(n+2))} \nonumber \\
	&=\sum_{n=0}^\infty\alpha(n+1)\int_0^1dx\int 
	\frac{d^{4} p}{(2 \pi)^{4}}\frac{p^2}{(p^2+\Delta'(x,n))^2},
\end{align}
where $\Delta'(x,n)=\alpha(n+2)-\alpha x$. 
Using \eqref{DR2} and performing the integral over $x$, we find 
\begin{align}
	A'&=-\frac{\alpha^2}{16\pi^2}\left(\frac{4\pi}{\alpha}\right)^\epsilon\Gamma(\epsilon-1)
	\sum_{n=0}^\infty(n+1)\left(\frac{1}{(n+1)^{\epsilon-2}}-\frac{1}{(n+2)^{\epsilon-2}}\right) 
	\nonumber \\
	&=-\frac{\alpha^2}{16\pi^2}\left(\frac{4\pi}{\alpha}\right)^\epsilon
	\Gamma(\epsilon-1)\Big(\zeta[\epsilon-3,1]-\zeta[\epsilon-3,2]+\zeta[\epsilon-2,2]\Big) 
	\nonumber \\
	&=-\frac{\alpha^2}{16\pi^2}\left(\frac{4\pi}{\alpha}\right)^\epsilon \Gamma(\epsilon-1) \zeta[\epsilon-2,0]. 
\end{align}
Thus, $I_{A3}$ is computed as
\begin{align}
	I_{A3}&\equiv I^{(2)}_{A3}+ I^{(3)}_{A3}=-\frac{32ig^2}{\Lambda^2}|N|(A+A') \nonumber \\
	&=+i\frac{4g^2\alpha^2|N|}{\pi^2 \Lambda^2}\left(\frac{4\pi}{\alpha}\right)^\epsilon 
	\Gamma(\epsilon-1) \zeta[\epsilon-2,0].
\end{align}

\subsection{Case \eqref{A4(2)} and \eqref{A4(3)}} \label{appA4}
To calculate \eqref{A4(2)}, $B$ is defined as
\begin{align}
	B&\equiv\sum_{n=0}^\infty\int\frac{d^4p}{(2\pi)^4}
	\frac{\alpha(n+1)\alpha\left(n+\frac{1}{2}\right)}{(p^2+\alpha n)(p^2+\alpha(n+1))} 
	\nonumber \\
	&=\alpha^2\sum_{n=0}^\infty(n+1)\left(n+\frac{1}{2}\right)
	\int_0^1dx\int\frac{d^4p}{(2\pi)^4}\frac{1}{(p^2+\Delta(x,n))^2},
\end{align}
where $\Delta(x,n)=\alpha(n+1)-\alpha x$.
Using \eqref{DR1}, $B$ is expressed as
\begin{align}
	B&=\alpha^2\sum_{n=0}^\infty(n+1)\left(n+\frac{1}{2}\right)\int_0^1dx\frac{1}{(4 \pi)^{d / 2}} 
	\Gamma\left(2-\frac{d}{2}\right)\left(\frac{1}{\Delta}\right)^{2-\frac{d}{2}} 
	\nonumber \\
	&=(-1)^{\epsilon}\frac{\alpha^2}{16\pi^2}\left(\frac{4\pi}{\alpha}\right)^\epsilon\Gamma(\epsilon)
	\sum_{n=0}^\infty(n+1)\left(n+\frac{1}{2}\right)\int_0^1dx\left(\frac{1}{x-(n+1)}\right)^{\epsilon}.
\end{align}
Performing the integral over $x$, we find
\begin{align}
	B&=-\frac{\alpha^2}{16\pi^2}\left(\frac{4\pi}{\alpha}\right)^\epsilon\Gamma(\epsilon-1)
	\sum_{n=0}^\infty(n+1)\left(n+\frac{1}{2}\right)\left(-\frac{1}{n^{\epsilon-1}}+\frac{1}{(n+1)^{\epsilon-1}}\right).
\end{align}
We compute the part of mode summation as follows. 
\begin{align*}
	(Sum)&=\sum_{n=0}^\infty(n+1)\left(n+\frac{1}{2}\right)
	\left(-\frac{1}{n^{\epsilon-1}}+\frac{1}{(n+1)^{\epsilon-1}}\right) \\
	&=\sum_{n=0}^\infty\left(-\frac{1}{n^{\epsilon-3}}-\frac{1}{2}
	\frac{1}{n^{\epsilon-2}}-\frac{1}{n^{\epsilon-2}}-\frac{1}{2}\frac{1}{n^{\epsilon-1}}
	+\frac{1}{(n+1)^{\epsilon-3}}-\frac{1}{2}\frac{1}{(n+1)^{\epsilon-2}}\right) \\
	&=-2\zeta[\epsilon-2,0]-\frac{1}{2}\zeta[\epsilon-1,0],
\end{align*}
where we use the Hurwitz zeta functions in \ref{LFZF}.
Thus, $B$ is computed as
\begin{align}
	B&=\frac{\alpha^2}{16\pi^2}\left(\frac{4\pi}{\alpha}\right)^\epsilon\Gamma(\epsilon-1)
	\left(2\zeta[\epsilon-2,0]+\frac{1}{2}\zeta[\epsilon-1,0]\right).
	\label{MCB1}
\end{align}
Similarly, to calculate \eqref{A4(3)}, $B'$ is defined as 
\begin{align}
	B'&\equiv\sum_{n=0}^\infty\int\frac{d^4p}{(2\pi)^4}
	\frac{\alpha(n+1)\alpha\left(n+\frac{3}{2}\right)}{(p^2+\alpha (n+1))(p^2+\alpha(n+2))} \nonumber \\
	&=\alpha^2\sum_{n=0}^\infty(n+1)\left(n+\frac{3}{2}\right)
	\int_0^1dx\int\frac{d^4p}{(2\pi)^4}\frac{1}{(p^2+\Delta'(x,n))^2},
\end{align}
where $\Delta'(x,n)=\alpha(n+2)-\alpha x$.
Using \eqref{DR1} and performing the integral over $x$,
\begin{align}
	B'&=-\frac{\alpha^2}{16\pi^2}\left(\frac{4\pi}{\alpha}\right)^\epsilon\Gamma(\epsilon-1)
	\sum_{n=0}^\infty(n+1)\left(n+\frac{3}{2}\right)\left(-\frac{1}{(n+1)^{\epsilon-1}}+\frac{1}{(n+2)^{\epsilon-1}}\right).
\end{align}
We compute the part of mode summation as was done in the previous case. 
\begin{align*}
	(Sum)&=\sum_{n=0}^\infty(n+1)\left(n+\frac{3}{2}\right)
	\left(-\frac{1}{(n+1)^{\epsilon-1}}+\frac{1}{(n+2)^{\epsilon-1}}\right) \\
	&=\sum_{n=0}^\infty\left(-\frac{1}{(n+1)^{\epsilon-3}}-\frac{1}{2}\frac{1}{(n+1)^{\epsilon-2}}
	+\frac{1}{(n+2)^{\epsilon-3}}-\frac{1}{2}\frac{1}{(n+2)^{\epsilon-2}} 
	\right. \nonumber \\
	&\qquad \left. -\frac{1}{(n+2)^{\epsilon-2}}+\frac{1}{2}\frac{1}{(n+2)^{\epsilon-1}}\right) \\
	&=-2\zeta[\epsilon-2,0]+\frac{1}{2}\zeta[\epsilon-1,0],
\end{align*}
where we use the value of Hurwitz zeta function in \ref{LFZF}.
Thus, $B'$ is computed as
\begin{align}
	B'&=\frac{\alpha^2}{16\pi^2}\left(\frac{4\pi}{\alpha}\right)^\epsilon\Gamma(\epsilon-1)
	\Big(2\zeta[\epsilon-2,0]-\frac{1}{2}\zeta[\epsilon-1,0]\Big).
	\label{MCB2}
\end{align}
Summing \eqref{MCB1} and \eqref{MCB2}, $I_{A4}$ is found to be
\begin{align}
	I_{A4}&\equiv I^{(2)}_{A4}+I^{(3)}_{A4}=-\frac{32ig^2}{\Lambda^2}|N|(B+B') \nonumber \\
	&=-i\frac{8g^2\alpha^2|N|}{\pi^2 \Lambda^2}\left(\frac{4\pi}{\alpha}\right)^\epsilon\Gamma(\epsilon-1)\zeta[\epsilon-2,0].
\end{align}

\subsection{Case \eqref{phi3(2)} and \eqref{phi3(3)}} \label{appphi3}
To calculate \eqref{phi3(2)} and \eqref{phi3(3)}, $C$ is defined as
\begin{align}
	C&\equiv\sum_{n=0}^{\infty} \int \frac{d^{4} p}{(2 \pi)^{4}} 
	\frac{\alpha(n+1) p^{2}}{\left(p^{2}+\alpha\left( n+\frac{1}{2}\right)\right)\left(p^{2}
	+\alpha\left(n+\frac{3}{2}\right)\right)} \nonumber \\
	&=\sum_{n=0}^{\infty}\alpha(n+1)\int_0^1dx \int \frac{d^{4} p}{(2 \pi)^{4}} \frac{p^{2}}{(p^{2}+\Delta(x,n))^2},
\end{align}
where $\Delta(x,n)=\alpha(n+3/2)-\alpha x$.
Using \eqref{DR2}, $C$ is expressed as 
\begin{align}
	C=(-1)^{\epsilon-1}\frac{\alpha^2}{16\pi^2}\left(\frac{4\pi}{\epsilon}\right)^\epsilon(2-\epsilon)\Gamma(\epsilon-1)
	\sum_{n=0}^\infty(n+1)\int_0^1dx\left(\frac{1}{x-\alpha\left(n+\frac{3}{2}\right)}\right)^{\epsilon-1}.
\end{align}
Performing the integral over $x$, we find 
\begin{align}
	C&=-\frac{\alpha^2}{16\pi^2}\left(\frac{4\pi}{\epsilon}\right)^\epsilon
	\Gamma(\epsilon-1)\sum_{n=0}^\infty(n+1)\left(\frac{1}{\left(n+\frac{1}{2}\right)^{\epsilon-2}}
	-\frac{1}{\left(n+\frac{3}{2}\right)^{\epsilon-2}}\right).
\end{align}
The part of mode summation is computed as follows. 
\begin{align*}
	(Sum)&=\sum_{n=0}^\infty(n+1)\left(\frac{1}{\left(n+\frac{1}{2}\right)^{\epsilon-2}}
	-\frac{1}{\left(n+\frac{3}{2}\right)^{\epsilon-2}}\right) \\
	&=\zeta[\epsilon-3,1/2]+\frac{1}{2}\zeta[\epsilon-2,1/2]
	-\zeta[\epsilon-3,3/2]+\frac{1}{2}\zeta[\epsilon-2,3/2] \\
	&=\zeta[\epsilon-2,1/2].
\end{align*}
Thus, $C$ is computed as
\begin{align*}
	C&=-\frac{\alpha^2}{16\pi^2}\left(\frac{4\pi}{\epsilon}\right)^\epsilon\Gamma(\epsilon-1)\zeta[\epsilon-2,1/2],
\end{align*}
and $I_{\varphi3}$ is expressed as
\begin{align}
	I_{\varphi3}&\equiv I^{(2)}_{\varphi3}+I^{(3)}_{\varphi3}
	=- \frac{8ig^2}{\Lambda^2}|N|C\times2 \nonumber \\
	&=+i\frac{g^2\alpha^2|N|}{\pi^2 \Lambda^2}\left(\frac{4\pi}{\epsilon}\right)^\epsilon\Gamma(\epsilon-1)\zeta[\epsilon-2,1/2]. 
\end{align}

\subsection{Case \eqref{phi4(2)} and \eqref{phi4(3)}} \label{appphi4}
To calculate \eqref{phi4(2)}, $D$ is defined as
\begin{align}
	D&\equiv\sum_{n=0}^{\infty} \int \frac{d^{4} p}{(2 \pi)^{4}} 
	\frac{\alpha(n+1)\alpha\left(n-\frac{1}{4}\right)}{\left(p^{2}+\alpha\left( n+\frac{1}{2}\right)\right)
	\left(p^{2}+\alpha\left(n+\frac{3}{2}\right)\right)} \nonumber \\
	&=\alpha^2\sum_{n=0}^\infty(n+1)\left(n-\frac{1}{4}\right)
	\int_0^1dx\int\frac{d^4p}{(2\pi)^4}\frac{1}{(p^2+\Delta(x,n))^2},
\end{align}
where $\Delta(x,n)=\alpha(n+3/2)-\alpha x$.
Using \eqref{DR1}, $D$ is expressed as
\begin{align}
	D&=(-1)^{\epsilon}\frac{\alpha^2}{16\pi^2}\left(\frac{4\pi}{\alpha}\right)^\epsilon
	\Gamma(\epsilon)\sum_{n=0}^\infty(n+1)\left(n-\frac{1}{4}\right)
	\int_0^1dx\left(\frac{1}{x-\left(n+\frac{3}{2}\right)}\right)^{\epsilon}.
\end{align}
Performing the integral over $x$, we find 
\begin{align}
	D&=-\frac{\alpha^2}{16\pi^2}\left(\frac{4\pi}{\alpha}\right)^\epsilon
	\Gamma(\epsilon-1)\sum_{n=0}^\infty(n+1)\left(n-\frac{1}{4}\right)
	\left(-\frac{1}{\left(n+\frac{1}{2}\right)^{\epsilon-1}}+\frac{1}{\left(n+\frac{3}{2}\right)^{\epsilon-1}}\right).
\end{align}
The part of mode summation is computed as
\begin{align*}
	(Sum)&=\sum_{n=0}^\infty(n+1)\left(n-\frac{1}{4}\right)
	\left(-\frac{1}{\left(n+\frac{1}{2}\right)^{\epsilon-1}}+\frac{1}{\left(n+\frac{3}{2}\right)^{\epsilon-1}}\right) \\
	&=-\zeta[\epsilon-3,1/2]+\frac{1}{4}\zeta[\epsilon-2,1/2]+\frac{3}{8}\zeta[\epsilon-1,1/2]
	+\zeta[\epsilon-3,3/2] \nonumber \\
	& \quad -\frac{9}{4}\zeta[\epsilon-2,3/2]+\frac{7}{8}\zeta[\epsilon-1,3/2] \\
	&=-2\zeta[\epsilon-2,1/2]+\frac{5}{4}\zeta[\epsilon-1,1/2],
\end{align*}
where we use the value of Hurwitz zeta function in \ref{LFZF}.
Thus, $D$ is computed as
\begin{align}
	D&=\frac{\alpha^2}{16\pi^2}\left(\frac{4\pi}{\alpha}\right)^\epsilon\Gamma(\epsilon-1)
	\Big(2\zeta[\epsilon-2,1/2]-\frac{5}{4}\zeta[\epsilon-1,1/2]\Big).
	\label{MCD1}
\end{align}
Similarly, to calculate \eqref{phi4(3)}, $D'$ is defined as 
\begin{align}
	D'&\equiv\sum_{n=0}^{\infty} \int \frac{d^{4} p}{(2 \pi)^{4}} 
	\frac{\alpha(n+1)\alpha\left(n+\frac{9}{4}\right)}{\left(p^{2}+\alpha\left( n+\frac{1}{2}\right)\right)
	\left(p^{2}+\alpha\left(n+\frac{3}{2}\right)\right)} \nonumber \\
	&=\alpha^2\sum_{n=0}^\infty(n+1)\left(n+\frac{9}{4}\right)
	\int_0^1dx\int\frac{d^4p}{(2\pi)^4}\frac{1}{(p^2+\Delta(x,n))^2},
\end{align}
where $\Delta(x,n)=\alpha(n+3/2)-\alpha x$.
Using \eqref{DR1} and performing the integral over $x$, we find 
\begin{align}
	D'=-\frac{\alpha^2}{16\pi^2}\left(\frac{4\pi}{\alpha}\right)^\epsilon\Gamma(\epsilon-1)
	\sum_{n=0}^\infty(n+1)\left(n+\frac{9}{4}\right)
	\left(-\frac{1}{\left(n+\frac{1}{2}\right)^{\epsilon-1}}+\frac{1}{\left(n+\frac{3}{2}\right)^{\epsilon-1}}\right).
\end{align}
The part of mode summation is computed as
\begin{align*}
	(Sum)&=\sum_{n=0}^\infty(n+1)\left(n+\frac{9}{4}\right)
	\left(-\frac{1}{\left(n+\frac{1}{2}\right)^{\epsilon-1}}+\frac{1}{\left(n+\frac{3}{2}\right)^{\epsilon-1}}\right) \\
	&=-\zeta[\epsilon-3,1/2]-\frac{9}{4}\zeta[\epsilon-2,1/2]-\frac{7}{8}\zeta[\epsilon-1,1/2]
	+\zeta[\epsilon-3,3/2] \nonumber \\
	& \quad +\frac{1}{4}\zeta[\epsilon-2,3/2]-\frac{3}{8}\zeta[\epsilon-1,3/2] \\
	&=-2\zeta[\epsilon-2,1/2]-\frac{5}{4}\zeta[\epsilon-1,1/2].
\end{align*}
where we use the Hurwitz zeta functions in \ref{LFZF}.
Thus, $D'$ is computed as
\begin{align}
	D'&=\frac{\alpha^2}{16\pi^2}\left(\frac{4\pi}{\alpha}\right)^\epsilon
	\Gamma(\epsilon-1)\Big(2\zeta[\epsilon-2,1/2]+\frac{5}{4}\zeta[\epsilon-1,1/2]\Big). 
	\label{MCD2}
\end{align}
Summing \eqref{MCD1} and \eqref{MCD2}, $I_{\varphi4}$ is expressed as
\begin{align}
	I_{\varphi4}&=I^{(2)}_{\varphi4}+I^{(3)}_{\varphi4}
	=-\frac{8ig^2}{\Lambda^2}|N|(D+D') \nonumber \\
	&=-i\frac{2g^2\alpha^2|N|}{\pi^2 \Lambda^2}
	\left(\frac{4\pi}{\alpha}\right)^\epsilon\Gamma(\epsilon-1)\zeta[\epsilon-2,1/2].
\end{align}


\end{document}